\documentclass[12pt]{article}
\usepackage{color} %can't use package color with jrl_thesis
\usepackage{graphicx}
\usepackage{fancyheadings}
\usepackage{natbib}
\usepackage{setspace}
\usepackage{amsmath,amssymb}
\usepackage{lastpage}
\usepackage{psfrag}
\usepackage{caption}
\usepackage{epsfig}
\usepackage{subfigure}
\newcommand{\Dpartial}[2]{ \frac{\partial #1}{\partial #2} }
\newcommand{\prfrac}[1]{\frac{\partial #1}{\partial r}}
\DeclareMathOperator{\Erf}{Erf}
\DeclareMathOperator{\sgn}{sgn}
\def\RR{{\mathbb{R}}}
\def\e{{\bf e}}

\def\x{{\bf x}}

\def\v{{\bf v}}
\def\u{{\bf u}}
\def\0{{\bf 0}}
\def\O{{\mathcal O}}
\def\F{{\bf F}}

\def\R{{\mathcal R}}

\def\I{{\mathcal I}}

\def\bA{{\bf A}}
\def\bB{{\bf B}}
\def\bF{{\bf F}}
\def\bR{{\bf R}}
\def\bW{{\bf W}}
\def\bX{{\bf X}}
\def\bnabla{\boldsymbol{\nabla}}

\def\tg{{\tilde{g}}}

\def\Dpartial#1#2{ {\partial #1 \over \partial #2} }

\def\Dpartialn#1#2#3{ {\partial^{#3} #1 \over \partial #2^{#3}} }

\newcommand{\diag}{\operatorname{diag}}

\newcommand{\plotstream}[1]{
\includegraphics[width=0.47 \textwidth]{Figs2/velocityRe#1.eps}}
\newcommand{\plotvorticity}[1]{
\includegraphics[width=0.47 \textwidth]{Figs2/vorticityRe#1.eps}}

\begin{document}

\title{Computation of Steady Incompressible Flows in Unbounded Domains}
\author{Jonathan Gustafsson$^{1,2}$ and Bartosz Protas$^{3,}$\thanks{Email address for correspondence: bprotas@mcmaster.ca} \\ \\
$^1$Center for Decision, Risk, Controls \& Signals Intelligence \\
Naval Postgraduate School, 93943 Monterey, USA \\ \\
$^2$School of Computational Science and Engineering \\ 
McMaster University, L8S 4K1 Hamilton, Canada \\ \\
$^3$Department of Mathematics and Statistics \\
McMaster University, Hamilton, ON, Canada
}

\maketitle
\begin{abstract}
  In this study we revisit the problem of computing steady
  Navier-Stokes flows in two-dimensional unbounded domains.  Precise
  quantitative characterization of such flows in the high-Reynolds
  number limit remains an open problem of theoretical fluid dynamics.
  Following a review of key mathematical properties of such solutions
  related to the slow decay of the velocity field at large distances
  from the obstacle, we develop and carefully validate a
  spectrally-accurate computational approach which ensures the correct
  behavior of the solution at infinity. In the proposed method the
  numerical solution is defined on the entire unbounded domain without
  the need to truncate this domain to a finite box with some
  artificial boundary conditions prescribed at its boundaries. Since
  our approach relies on the streamfunction-vorticity formulation, the
  main complication is the presence of a discontinuity in the
  streamfunction field at infinity which is related to the slow decay
  of this field. We demonstrate how this difficulty can be overcome by
  reformulating the problem using a suitable background ''skeleton''
  field expressed in terms of the corresponding Oseen flow combined
  with spectral filtering.  The method is thoroughly validated for
  Reynolds numbers spanning two orders of magnitude with the results
  comparing favourably against known theoretical predictions and the
  data available in the literature.
\end{abstract}
%\begin{keyword}
\begin{flushleft}
Keywords:
Steady Navier-Stokes system; unbounded domains; wake flows; spectral methods
\end{flushleft}
%\end{keyword}

%\tableofcontents

\section{Introduction}
\label{sec:intro}

In this work we revisit the classical problem of computing steady
flows past an obstacle in an unbounded domain which has played an
important role in theoretical fluid mechanics, especially, in the
study of separated flows \cite{srsk98}. An aspect of this problem
which has received particular attention is the structure of the flow
field in the limit when the Reynolds number $Re \rightarrow \infty$.
It is well known that the inviscid Euler flows in the same geometric
setting admit several different solutions with quite distinct
properties --- in addition to the Kirchhoff free-streamline flows
featuring an open wake region extending to infinity
\cite{lc07,Brodetsky1923}, flows with compact vorticity regions
predicted by the Prandtl-Batchelor theory \cite{Batchelor1956} have
also been found \cite{Elcrat2000}. Perturbation-type solutions to this
problem were constructed using methods of asymptotic analysis by
Chernyshenko \cite{ch88,ch98a}. While these solutions remain the most
advanced theoretical results concerning this problem, their
computational validation for large $Re$ remains an open problem with
Fornberg's results from the late 1980s still representing the
state-of-the-art \cite{Fornberg1980,Fornberg1985}. As will be argued
below, what makes this problem challenging from the computational
point of view is the combination of steadiness and an unbounded domain
which results in a very slow decay of the flow fields towards their
limiting values at large distances from the obstacle. In the recent
years significant advances have been made as regards mathematical
characterization of such flows \cite{Galdi2011}, and the goal of this
work is to develop and validate a numerical approach which explicitly
accounts for these properties. More specifically, the proposed
technique will achieve the spectral accuracy for solutions defined on
unbounded domains (i.e., without the need to truncate the domain to a
finite ``computational box'' with some artificial boundary conditions
prescribed on its boundaries) and will in addition ensure that
solutions have the right asymptotic behavior at large distances from
the obstacle.

\begin{figure}
\begin{center}
\includegraphics[width=0.9\textwidth]{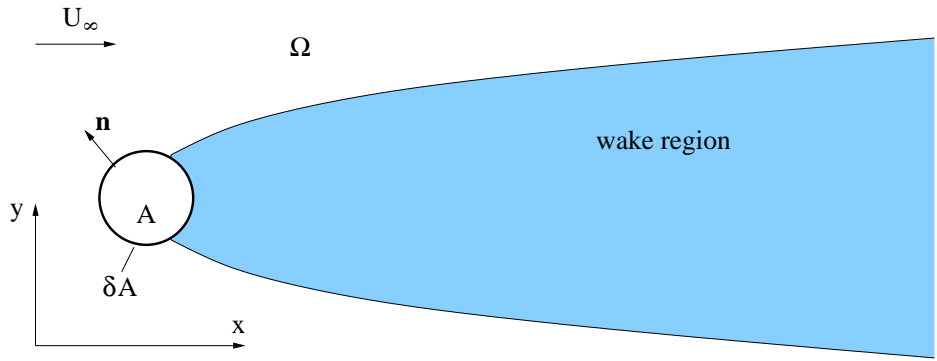}
\caption{Geometry of the flow domain $\Omega$ with a schematic
  representation of the wake region (shaded) characterized by the slow
  decay of the flow field to its asymptotic values.}
\label{fig:wake}
\end{center}
\end{figure}

We thus consider the problem defined on the two-dimensional (2D)
unbounded domain $\Omega$ which is the exterior of a circular obstacle
$A$ of diameter $d$ (Figure \ref{fig:wake}). Given the free stream
velocity at infinity $U_{\infty}$, the system of equations we are
interested in is
\begin{subequations}\label{eq:steadyNSBC}
\begin{alignat}{2}
({\bf v} \cdot \nabla) {\bf v} & = - \nabla p+\frac{1}{Re}\Delta {\bf v} \qquad && \textrm{in} \ \Omega, \label{eq:steadyNSBCmomentum}\\
\nabla \cdot {\bf v} & = 0 && \textrm{in} \ \Omega, \label{eq:steadyNSBCincomp}\\
{\bf v} & = {\bf 0} && \textrm{on} \ \partial A,\label{eq:steadyNSBCc}\\
{\bf v} & \rightarrow U_{\infty}\e_x  && \textrm{as} \ |\x| \rightarrow \infty, \label{eq:steadyNSBCd}
\end{alignat}
\end{subequations}
where ${\bf v} = [u,v]$ is the velocity vector, $p$ is pressure,
$\e_x$ is the unit vector associated with the X-axis, $\x=[x,y] \in \Omega$
is the position vector and $Re := U_{\infty} d / \nu$ is the Reynolds
number in which $\nu$ is the kinematic viscosity (for simplicity, the
fluid density is set equal to one). The symbol ``$:=$'' means ``equal
to by definition''. Mathematical analysis of problem
\eqref{eq:steadyNSBC}, which was initiated by Leray in the 1930s
\cite{Leray1933} and continued by Finn in the 1960s
\cite{Finn1965,Finn1967a,Finn1967b}, is surveyed in the monograph
by Galdi \cite{Galdi2011}. It reveals a number of interesting
properties related to the behavior of the velocity field at
large distances from the obstacle which is quite distinct from the
corresponding time-dependent flows. More precisely, steady 2D flows
described by \eqref{eq:steadyNSBC} feature a ``wake'' region in the
direction of the X-axis, cf.~Figure \ref{fig:wake}, in which the
velocity field $\v$ approaches its asymptotic value $U_{\infty}\e_x$
much slower than outside this region, namely at the rate
\begin{equation}
|\v(\x) - U_{\infty}\e_x|  = \O(|\x|^{-1/4-\epsilon}) \qquad \textrm{as} \ |\x| \rightarrow \infty,
\label{eq:vinf}
\end{equation}
where $\epsilon>0$. Solutions of this type were referred to by Finn as
``physically reasonable'' (PR) \cite{Finn1965} and have the additional
property that to the leading order they have the same behavior at
large distances as the solutions of the corresponding Oseen problem
characterized by the same drag force \cite{Galdi2011}, i.e.,
\begin{equation}\label{eq:PRvel}
\v(\x) = U_{\infty}\e_x +{\bf F} \cdot {\bf E(x)}+ {\bf V(x)} \qquad \textrm{as} \ |\x| \rightarrow \infty,
\end{equation}
where $\F = [F_x \ F_y]^T$ is the hydrodynamic force acting on the
obstacle $A$, ${\bf E(x)}$ is the fundamental solution tensor for the
Oseen system
\begin{subequations}
\label{eq:oseen}
\begin{alignat}{2}
(U_{\infty}\e_x)\cdot\bnabla\u + \bnabla p  - \frac{1}{Re} \Delta \u  & =   \0  \qquad &  & \text{in} \ \Omega,
        \label{eq:oseen_a} \\
\bnabla \cdot \u  &  =  0  &  & \text{in} \ \Omega,
        \label{eq:oseen_b}\\
      \u & =  {\bf 0}  &  & \text{on} \  \partial A,
        \label{eq:oseen_c}\\ 
      \u &  \rightarrow  \u_{\infty}  &  & \text{as} \  |\x| \rightarrow\infty,
        \label{eq:oseen_d}
\end{alignat}
\end{subequations}
and the ``remainder'' ${\bf V(x)}$ satisfies the following asymptotic estimate
\begin{equation}
{\bf V(x)} = \O(|\x|^{-1}\log^2{|\x|})  \qquad \textrm{as} \ |\x| \rightarrow \infty.
\end{equation}
In other words, at large distances from the obstacle the PR solutions
are up to a rapidly vanishing correction indistinguishable from the
Oseen flows exhibiting the same drag $\F$. Finn and Smith
\cite{Finn1967a} showed that for small Reynolds numbers problem
\eqref{eq:steadyNSBC} has at least one solution that is physically
reasonable. While it remains to be proven whether steady
Navier-Stokes system \eqref{eq:steadyNSBC} has solutions for {\em
  all} values of the Reynolds number, for now we will assume that at
least one solution exists for all finite Reynolds numbers. In addition
to making the numerical solution of problem \eqref{eq:steadyNSBC} more
challenging, the properties discussed above also complicate evaluation
of the hydrodynamic forces \cite{p11}.

The first calculation of a steady flow around a circular cylinder was
carried out by Thom \cite{Thom1933} for low Reynolds numbers ($Re =
10-20$) using the streamfunction-vorticity formulation. An interesting
aspect of that research was the use of a conformal mapping.  The
simulations performed by Kawaguti \cite{Kawaguti1953} and by Apelt
\cite{Apelt1961} for the Reynolds number up to 44 showed a linear
growth of the vortex pair behind the cylinder with $Re$. Allen and
Southwell \cite{Allen1955} introduced upwind schemes to computational
fluid dynamics when solving steady flows for Reynolds numbers up to
1000.  Their solutions showed a trend of reduced recirculation length
for the Reynolds number increasing from 10 to 100. The results of
Hamielec and Raal \cite{Hamielec1969} also indicated that the
recirculation length decreased for Reynolds number larger than 50.  We
remark that, as discussed below, these results are now believed to be
erroneous. Keller \cite{Keller1966} and Takami \cite{Takami1969}
combined conformal mappings with finite-difference methods to solve
steady flows around the cylinder for the Reynolds number up to 15,
whereas a spectral method for the study of the stability of flows in
unbounded domains was developed by Zebib \cite{z87}.  These earlier
investigations are reviewed in the historical survey by Fornberg
\cite{Fornberg1993}.  Many numerical difficulties in solving system
\eqref{eq:steadyNSBC} stem from the fact that the unbounded domain
$\Omega$ needed to be truncated to a finite computational box and it
is not immediately obvious what boundary conditions must be prescribed
on its boundary to ensure the solutions exhibit the correct asymptotic
behavior given in \eqref{eq:vinf}--\eqref{eq:PRvel}.

\begin{figure}
\begin{center}
\mbox{
\subfigure[]{
\includegraphics[width=0.48\textwidth]{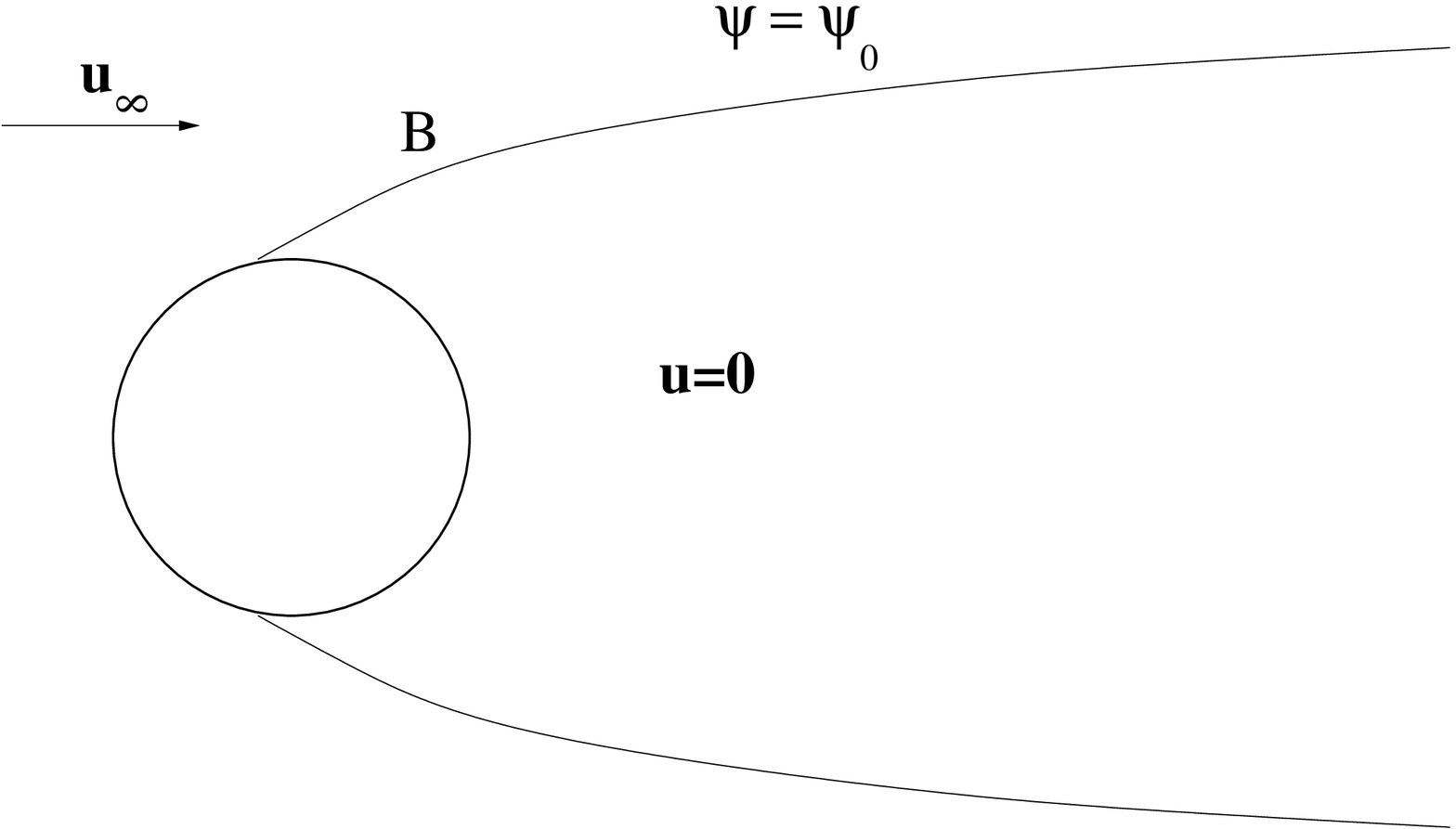}
}\qquad
\subfigure[]{
\includegraphics[width=0.48\textwidth]{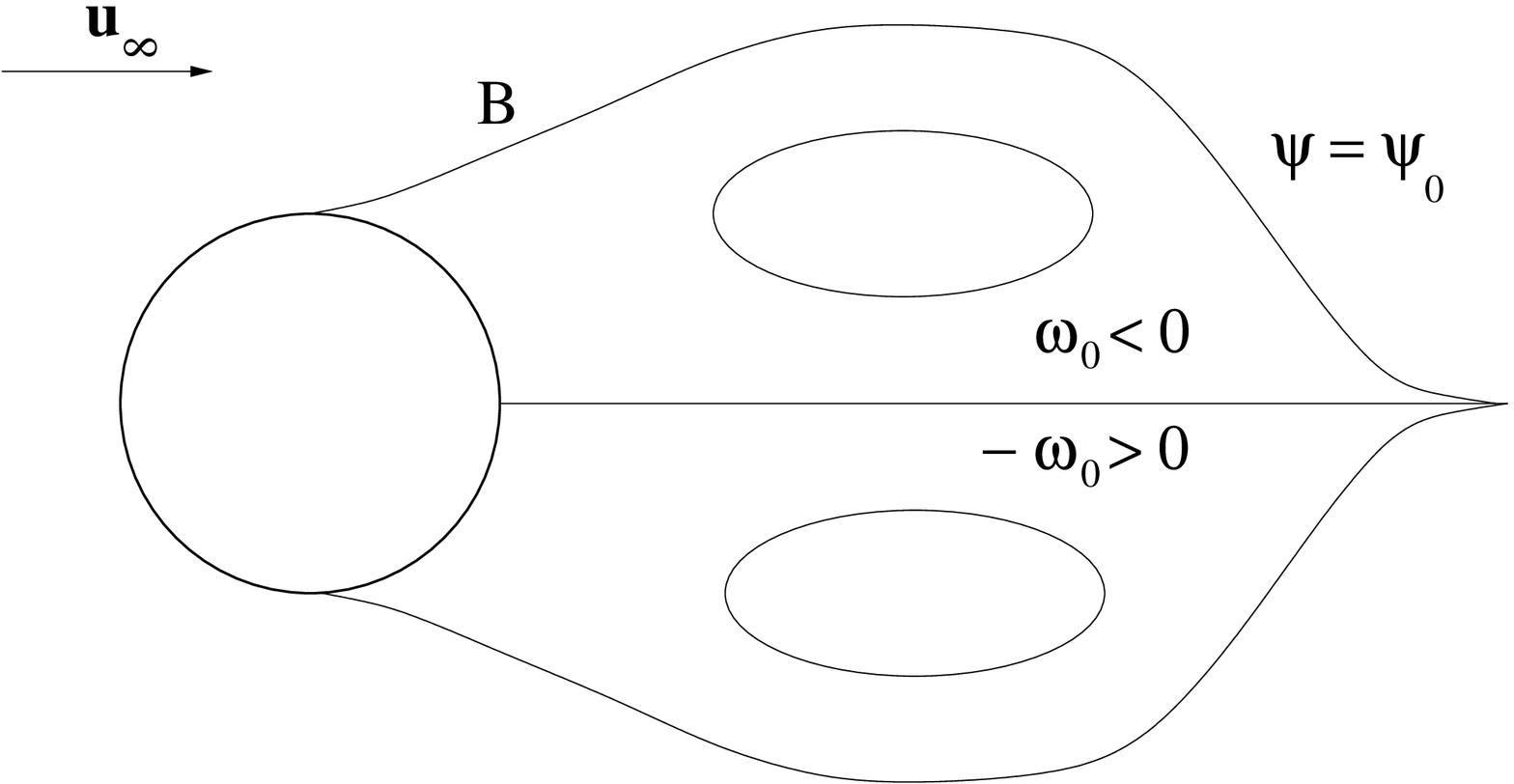}
}
}
\caption{Schematic showing the main features of the separation zone in
  (a) Kirchhoff's model \cite{Kirchoff1869,lc07} and (b) Batchelor's
  model \cite{Batchelor1956} of the steady wake flow in the infinite
  Reynolds number limit.}
\label{fig:infRe}
\end{center}
\end{figure}

The significance of the far-field boundary conditions was already
recognized by Fornberg \cite{Fornberg1985} who observed that the use
of the free-stream values on the outer boundary of the computational
domain produced large errors even for low Reynolds number. We note,
however, that Fornberg considered the free-stream values for the
streamfunction only while setting the vorticity equal to zero. In the
numerical results of Fornberg \cite{Fornberg1985} the length of the
recirculation zone appears proportional to the Reynolds number. The
recirculation width, however, exhibits different behaviour depending
on the Reynolds number: for $Re\lessapprox 300$ the width appears
proportional to the square root of the Reynolds number; on the other
hand, for $Re \gtrapprox 300$ the relation is linear. This behaviour
is also reflected in the different flow patterns observed in the two
regimes with the flows obtained for $Re\lessapprox 300$ featuring a
slender wake reminiscent of the Kirchhoff free-streamline solution
(\cite{Kirchoff1869,lc07}, see Figure \ref{fig:infRe}a) and those
corresponding to $Re \gtrapprox 300$ characterized by a wider
recirculation region more similar to the Prandtl-Batchelor limiting
solution (\cite{Batchelor1956}, see Figure \ref{fig:infRe}b). Thus,
although Fornberg's solutions \cite{Fornberg1985} still represent the
state-of-the-art in this field, they are rather inconclusive as
regards the solution structure at large distances in the high-Reynolds
number limit. There exist more recent results concerning two
dimensional steady-state flows past obstacles, but they involve
different configurations such as flows past arrays of obstacles as in
\cite{Fornberg1991,ga04}, flows past obstacles in channels
\cite{smb09}, or flows of stratified fluids \cite{cc96}.

The question of consistent boundary condition imposed on the
boundaries of the computational domain was recently taken up by
B\"onisch et al. \cite{bhw05,Bonisch2006,bhw07}. As will be discussed
below, they devised an adaptive approach in which the corrections to
the free-stream are consistent with \eqref{eq:PRvel} and depend on the
force experienced by the obstacle. More recent attempts at solving
problem \eqref{eq:steadyNSBC}, although not necessarily focusing on
obtaining solutions in the high-$Re$ limit, include
\cite{v09,cmm09,gbl13} with study \cite{gbl13} containing certain
similar ideas to those investigated here. In the context of
time-dependent flows, the question of surrogate boundary conditions on
truncated domains was recently also addressed in \cite{dkc14}.

The main contribution of our study is development of a
spectrally-accurate solution method based on the
streamfunction-vorticity formulation ensuring that asymptotic
condition \eqref{eq:PRvel} is satisfied. As discussed below, the key
technical difficulty in this approach is the resolution of the
singularity appearing at infinity in the streamfunction field which is
achieved through a suitable change of the dependent variables together
with spectral filtering. To the best of our knowledge, this is the
first time this issue is addressed in the CFD literature.  The plan of
the paper is as follows: in the next Section we describe how the
steady Navier-Stokes system \eqref{eq:steadyNSBC} can be reformulated
as a suitable perturbation to Oseen system \eqref{eq:oseen}; in
Section \ref{sec:numer} we introduce key elements of the proposed
numerical approach; validation and computational results are presented
in Section \ref{sec:results}, whereas summary and conclusions are
deferred to Section \ref{sec:final}. For completeness, some technical
material is collected in Appendix.

\section{Steady Navier-Stokes Flows as Perturbations of Oseen Flows}
\label{sec:perturbation}

In this Section we introduce a transformation of system
\eqref{eq:steadyNSBC} which will allow us to enforce asymptotic
properties \eqref{eq:vinf}--\eqref{eq:PRvel} by construction in the
numerical solutions. As a point of departure, we transform system
\eqref{eq:steadyNSBC} to the frame of reference in which obstacle $A$
is moving with velocity $-U_{\infty}\e_x$ and there is no flow at
infinity. Expressing the solution in terms of streamfunction $\psi$
and vorticity $\omega = \Dpartial{u}{y}-\Dpartial{v}{x}$, we obtain
\begin{subequations}
\label{eq:streamvorticity}
\begin{alignat}{2}
({\bf v} \cdot \nabla ) \omega & = \frac{1}{Re} \Delta \omega  \qquad &&\textrm{in} \ \Omega, \label{eq:streamvorticitya}\\
\Delta \psi +\omega &=0 \qquad &&\textrm{in} \ \Omega, \label{eq:streamvorticityb} \\
{\bf v} & = -U_{\infty}\e_x && \textrm{on} \ \partial A, \\
{\bf v} &\rightarrow {\bf 0} && \textrm{as} \ |\x| \rightarrow \infty.
\end{alignat}
\end{subequations}
The streamfunction $\psi$ and velocity $\v$ are related as follows in,
respectively, the Cartesian $\{\e_x,\e_y\}$ and polar
$\{\e_r,\e_\theta\}$ coordinate systems
\begin{equation}
\v = \Dpartial{\psi}{y} \e_x - \Dpartial{\psi}{x} \e_y = \frac{1}{r}\Dpartial{\psi}{\theta} \e_r - \Dpartial{\psi}{r} \e_{\theta},
\end{equation}
where $r=|\x|$. When expressed in terms of the streamfunction, the
boundary conditions on the surface of the cylinder become
\begin{subequations}
\begin{align}\label{eq:BC}
\left. \psi(r,\theta) \right|_{r=1} &= -\sin{\theta}, & \theta \in{[0,\pi]},\\
\left. \frac{\partial \psi(r,\theta)}{\partial r} \right|_{r=1} &= -\sin{\theta},& \theta \in{[0,\pi]}, \label{eq:derivativeBC}
\end{align}
\end{subequations}
whereas for the boundary conditions at infinity we have
\begin{subequations}
\begin{align}
\lim_{r \rightarrow \infty} \frac{1}{r} \frac{\partial \psi(r,\theta)}{\partial \theta} &= 0,& \theta \in{[0,\pi]}, \\
\lim_{r \rightarrow \infty} -\frac{\partial \psi(r,\theta)}{\partial r} &= 0, & \theta \in{[0,\pi]}. \label{eq:streamBCb}
\end{align}
\end{subequations}
Equation \eqref{eq:streamBCb} implies that $\lim_{r \rightarrow
  \infty} \psi(r,\theta) = f(\theta)$, where $f(\theta)$ was shown by
Imai \cite{Imai1951} to be the leading-order term in the solution of
Oseen system \eqref{eq:oseen}. We note that, as already discussed in
\cite{Imai1951}, the leading-order term has a discontinuity with
respect to the azimuthal angle $\theta$ in the limit $r \rightarrow
\infty$, however, the associated velocity field is finite. It also has
the property that the corresponding vorticity inside the wake region,
characterized by $\theta = \O(r^{\frac{1}{2}})$, behaves as
$\O(r^{-\frac{1}{2}})$ when $r \rightarrow \infty$, whereas outside
this region the vorticity vanishes exponentially as $r \rightarrow
\infty$. As one possibility, the limiting function $f$ can be
therefore given by $f(\theta) = \lim_{r \rightarrow \infty}
g(r,\theta)$, where
\begin{equation}\label{eq:asymptoticstream}
g(r,\theta) = \frac{F_x}{2} \left[\frac{\theta}{\pi}-\mathcal{H}(\cos{\theta})\Erf \left( \sqrt{r} \sin{\left( \frac{\theta}{2} \right)} \right) \right]
\end{equation}
in which $\Erf (x) := \frac{2}{\sqrt{\pi}}\int_0^x e^{-t^2}\,dt$ and
$\mathcal{H}(x)$ is the Heaviside step function defined as
\begin{equation}\label{eq:Heaviside}
\mathcal{H}(x) := \ \begin{cases}
0, & x<0, \\
\frac{1}{2}, & x = 0, \\
1, & x >0.
\end{cases}
\end{equation}
The step function is needed in order to prevent an extra jump of the
streamfunction at $\theta= \pm\pi/2$. The drag force $F_x$ appearing in
\eqref{eq:asymptoticstream} can be numerically calculated using two
different methods. The first one is borrowed from \cite{Fornberg1980}
\begin{equation}\label{eq:dragFornberg}
F_x =  \frac{2}{Re} \int_0^{\pi}\left. \left( r \prfrac{\omega} - \omega \right) \right|_{\partial A} \sin{\theta} d \theta,
\end{equation}
whereas the second was proposed in \cite{Veysey2007}
\begin{equation}\label{eq:dragVeysey}
F_x =  \frac{2}{Re} \int_0^{\pi}\left. -r^2 \frac{\partial^3 \psi}{\partial r^3} \right|_{\partial A} \sin{\theta} d \theta.
\end{equation}

\begin{figure}\centering{
\includegraphics[width=0.6\textwidth]{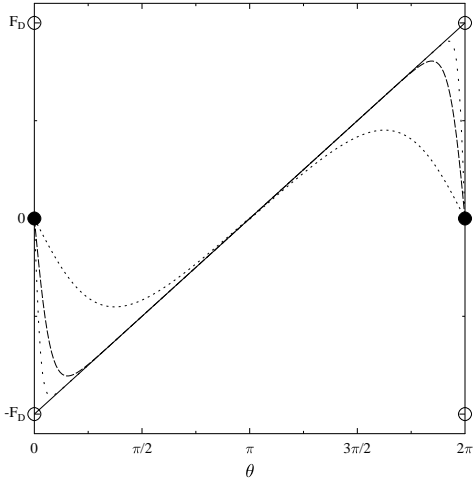}}
\caption{Dependence of the streamfunction $\psi$ on the azimuthal
  angle $\theta$ for increasing values of $r$ (represented by function
  $g(r,\theta)$ marked by the dotted and dashed lines) and in the
  limit as $r \rightarrow \infty$ (represented by function $f(\theta)$
  marked by the solid line with empty and solid symbols to denote the
  discontinuity).
  \label{fig:streaminf}}
\end{figure}

We note that function $g(r,\theta)$ as defined in
\eqref{eq:asymptoticstream} is continuous in $\theta$ for all {\it
  finite} r, but has a discontinuity at $\theta=0$ in the limit when
$r \rightarrow \infty$. In agreement with the results reviewed in
\cite{Galdi2011}, this discontinuity is a consequence of the
asymptotic behaviour for the velocity field inside the wake region. A
plot of functions $f(\theta)$ and $g(r,\theta)$ for several increasing
values of $r$ is shown in Figure \ref{fig:streaminf}.

The main problem from the numerical point of view is that the
streamfunction tends towards its asymptotic value far away from the
cylinder very slowly making domain truncation difficult and the
asymptotic value is not continuous in the azimuthal direction. To
remedy this problem, we will represent solutions to steady
Navier-Stokes system \eqref{eq:streamvorticity} in the original
unbounded domain $\Omega$ as perturbations to a suitable ``skeleton''
given analytically and designed to capture the asymptotic behaviour of
the Navier-Stokes flows at infinity,
cf.~\eqref{eq:vinf}--\eqref{eq:PRvel}. Consequently, the perturbation
field, which has to be obtained numerically, will be more localized
than the entire solution and hence easier to compute. More precisely,
we will represent the solution to system \eqref{eq:streamvorticity} as
\begin{equation}
\psi(r,\theta) = \psi'(r,\theta)+g(r,\theta), \qquad 
\omega(r,\theta) = \omega'(r,\theta)
\label{eq:psi'}
\end{equation}
in which $\psi'$ and $\omega'$ are the perturbation streamfunction and
vorticity, whereas the skeleton $g$ will be obtained as a
leading-order term in the far-field expansion of the solution to Oseen
system \eqref{eq:oseen}. We add that defining the ``skeleton'' in
terms of the streamfunction will automatically ensure its
incompressibility.  We will make the following assumption about
function $g$
\begin{equation}\label{eq:g}
{g}|_{r=1} = \left. \frac{\partial {g}}{\partial r} \right|_{r=1} 
= \left. \frac{\partial^2 {g}}{\partial^2 r} \right|_{r=1} 
= \left. \frac{\partial^3 {g}}{\partial^3 r} \right|_{r=1} =0.
\end{equation}
As a result, the boundary conditions satisfied by $\psi$ and $\psi'$
on the cylinder boundary $\partial A$ are the same and function $g$
does not affect the calculation of the drag. Possible choices of
function $g$ are discussed below in Sections \ref{sec:Wittver} and
\ref{sec:asymptoticoseen}. Although this function is constructed to
capture the flow structure at large distances from the obstacle, near
the obstacle it may exhibit a very different behaviour from the actual
flow solutions and a suitable ``mask'' function will be introduced in
Section \ref{sec:system} to compensate for this effect.  Combining
ansatz \eqref{eq:psi'} with Navier-Stokes system
\eqref{eq:streamvorticity}, we obtain the following system satisfied
by perturbation field $\psi'$ and the corresponding vorticity
$\omega'$ (for simplicity and with a slight abuse of notation, we will
hereafter drop the primes $(')$ and will use $\psi$ and $\omega$ in
lieu of $\psi'$ and $\omega'$)
\begin{subequations}\label{eq:streamvorticitybootstrap} 
\begin{multline} 
\frac{1}{r} \frac{\partial \psi}{\partial \theta} \frac{\partial \omega}{\partial r}  -\frac{1}{r}\frac{\partial \psi}{\partial r}\frac{\partial \omega}{\partial \theta} +\left( \frac{1}{r} \frac{\partial g}{\partial \theta} \prfrac{}{}-\frac{\partial g}{\partial r} \frac{1}{r}\frac{\partial}{\partial \theta} \right) \omega  \label{eq:streamvorticitybootstrapa} \\
-\frac{2}{Re}\left( \frac{\partial^2}{\partial r^2}+\frac{1}{r}\prfrac{}+\frac{1}{r^2} \frac{\partial^2}{\partial \theta^2} \right) \omega = 0,  \quad r \in{[1, \infty)},  \quad \theta \in{[0,\pi]},
\end{multline}
\begin{align}
\Delta \psi +\omega &=  -\Delta g, &&r \in{[1, \infty)}, &\theta \in{[0,\pi]}, \label{eq:streamvorticitybootstrapb} \\
\psi &= -g, &&r = 1, &\theta \in{[0,\pi]}, \label{eq:streamvorticitybootstrapc} \\
\frac{\partial \psi}{\partial r} &= -\frac{\partial g}{\partial r}, &&r = 1, &\theta \in{[0,\pi]}, \label{eq:streamvorticitybootstrapd} \\
\psi = f(\theta)&+r\sin{\theta}-g, &&\text{as $r \rightarrow \infty$}, &\theta \in{[0,\pi]}, \label{eq:streamvorticitybootstrape} \\
\omega &= 0, &&\text{as $r \rightarrow \infty$}, &\theta \in{[0,\pi]}. \label{eq:streamvorticitybootstrapf}
\end{align}
\end{subequations}
A spectral approach to the numerical solution of this problem in
unbounded domain $\Omega$ is discussed in Section \ref{sec:numer}.

We now discuss two different ways of constructing the skeleton
function $g(r,\theta)$, both of which are motivated by the analysis
of the solutions of Oseen problem \eqref{eq:oseen}, see also
\cite{Gustafsson2012}.

\subsection{Earlier Approaches}
\label{sec:Wittver}

B\"onisch et al.~\cite{bhw05} were interested in steady flows at
relatively low Reynolds numbers. While they used domain truncation,
the main novelty of their approach was a very careful choice of the
velocity boundary conditions prescribed on the boundary of the
computational domain. The following expressions were used
\begin{subequations}\label{eq:Wittvercartesian}
\begin{align}\label{eq:Wittverx}
u(x,y) &= F_x \left[ \frac{x}{\pi(x^2+y^2)}-\mathcal{H}(x)\frac{1}{\sqrt{\pi x }}e^{-\frac{y^2}{4x}}\right], \\
v(x,y) &= F_x \left[ \frac{y}{\pi(x^2+y^2)}-\mathcal{H}(x)\frac{y}{2\sqrt{\pi} x^{3/2}}e^{-\frac{y^2}{4x}}\right]
\end{align}
\end{subequations}
which are the leading-order terms of the solution describing the
incompressible flow around an obstacle of any shape when the Reynolds
number is low (the geometry of the obstacle is not important). As can
be verified, the longitudinal velocity component exhibits the expected
slow decay (proportional to $r^{-1/2}$) to its asymptotic value in the
wake region. In \cite{Bonisch2006} additional terms were introduced in
order to achieve a more accurate representation. Function $g$
corresponding to velocity field \eqref{eq:Wittvercartesian} takes the
following form in the polar coordinates
\begin{equation}\label{eq:Wittverstream}
g(r,\theta) = \frac{-F_x \,\arctan{\left( \frac{\cos{\theta}}{\sin{\theta}}\right)}}{\pi}-\frac{F_x}{2}
\begin{cases}
\Erf{\left( \frac{\sqrt{r}\sin{\theta}}{2\sqrt{\cos{\theta}}}\right)},& \textrm{if $\cos{\theta}>0$},\\ 
1, & \textrm{if $ \frac{\pi}{2} \leq \theta < \pi$}, \\
-1, & \textrm{if $ \pi < \theta \leq \frac{3\pi}{2}$}, \\
0, &\textrm{elsewhere}
\end{cases}
\end{equation}
and the corresponding vorticity field is
\begin{equation}\label{eq:Wittveromega}
\omega = - F_x \mathcal{H}(\cos{\theta})\frac{r \sin{\theta}}{2\sqrt{\pi}}
\left[6r\cos{\theta}-\frac{(r \sin{\theta})^2}{4 (r \cos{\theta})^{7/2}}-
\frac{1}{(r \cos{\theta})^{3/2}} \right] \, e^{-r\frac{\sin^2{\theta}}{4\cos{\theta}}}.
\end{equation}
We refer the reader to \cite{gPhD} for a comparison of the predictions
made based on expressions
\eqref{eq:Wittverstream}--\eqref{eq:Wittveromega} with the data
obtained from an actual solution of Navier-Stokes system
\eqref{eq:steadyNSBC} without ansatz \eqref{eq:psi'}.

\subsection{Present Approach}
\label{sec:asymptoticoseen}

\begin{figure}
\mbox{\hspace*{-0.5cm}
\subfigure[$g(r_0,\theta)$]{
\includegraphics[width=0.51\textwidth]{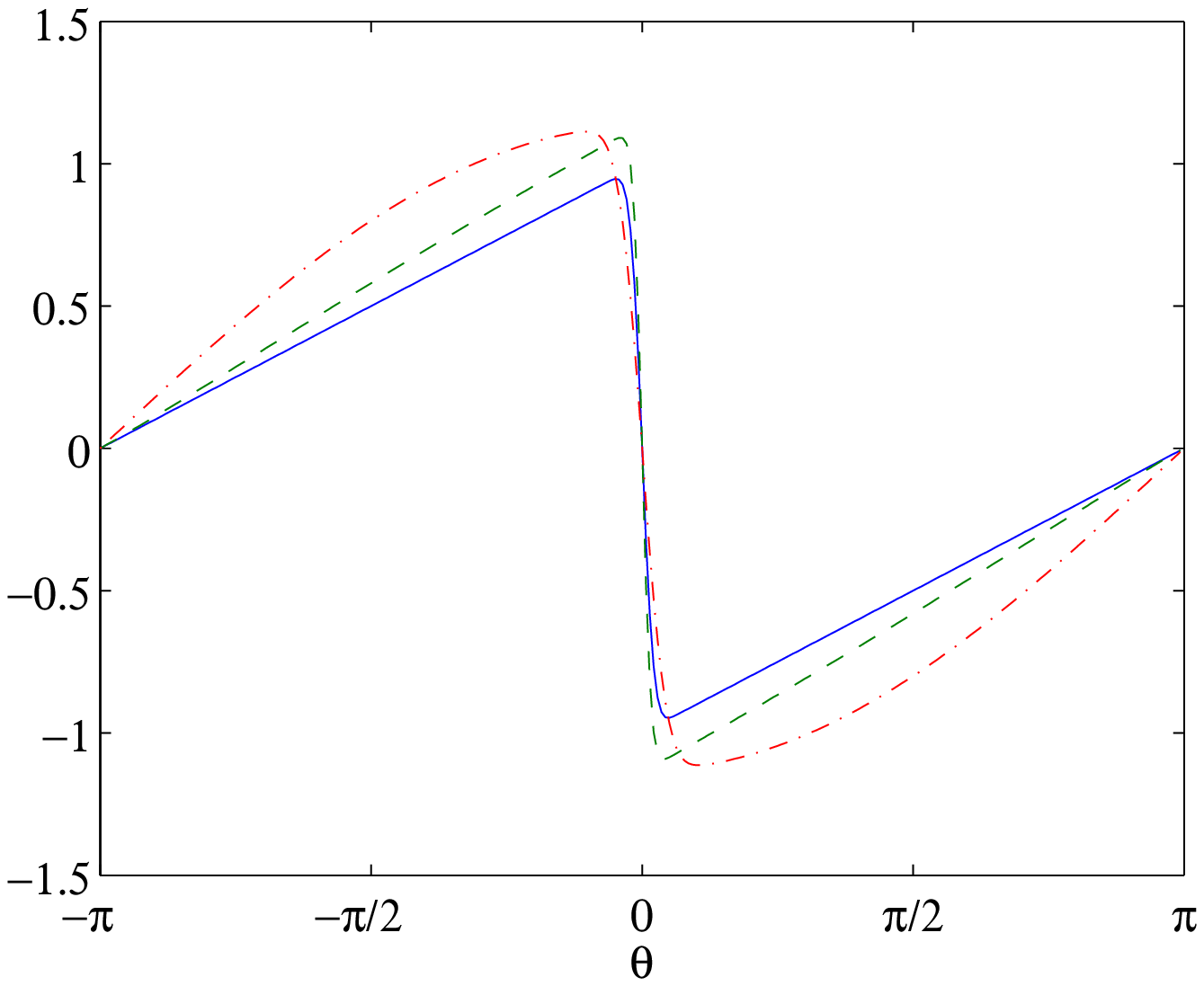}\label{fig:easystreamRe5}}
\subfigure[$-\Delta g(r_0,\theta)$]{
\includegraphics[width=0.51\textwidth]{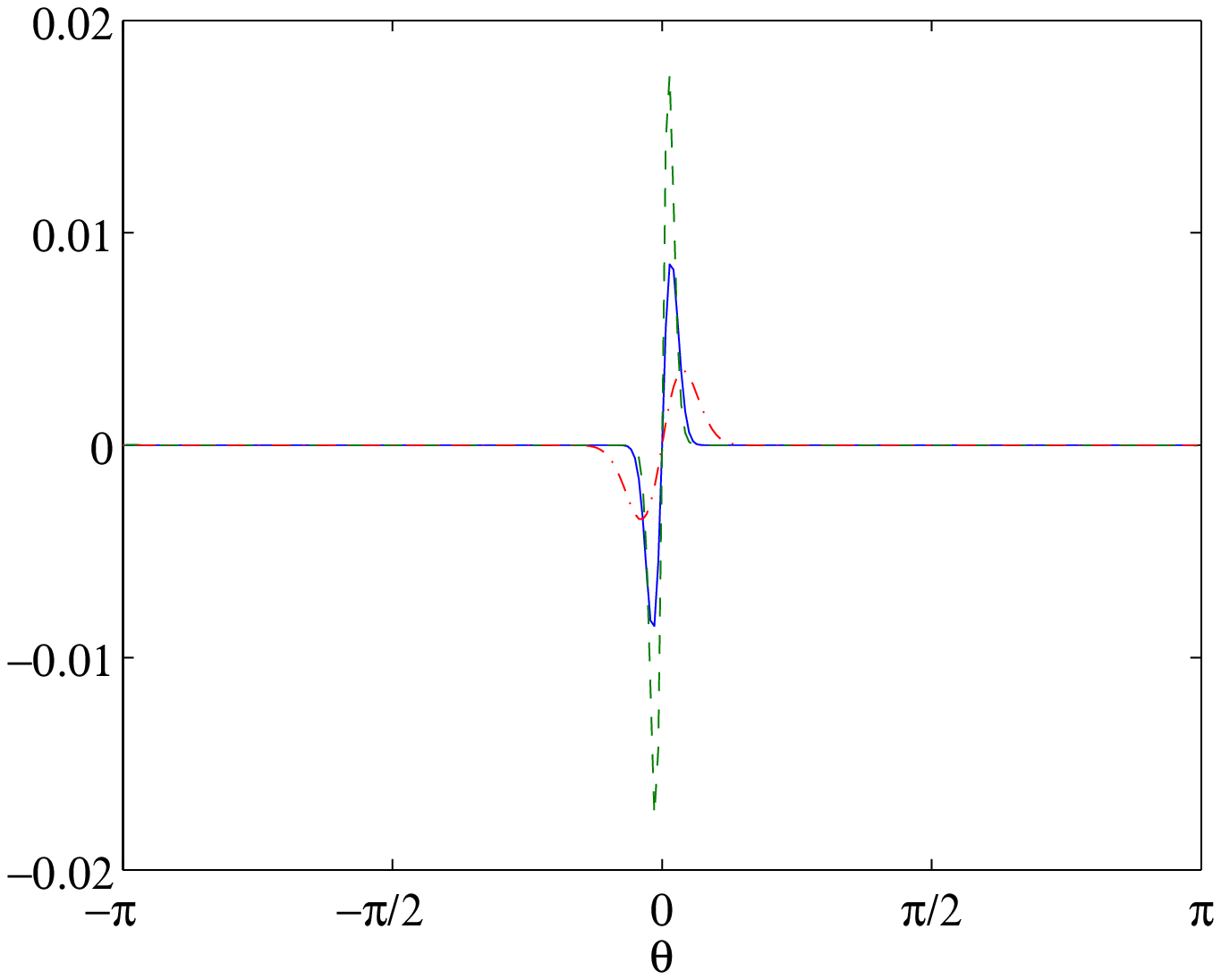}\label{fig:easyvorticityRe5}}}
\mbox{\hspace*{-0.5cm}
\subfigure[$\frac{1}{r_0}\Dpartial{g}{\theta}(r_0,\theta)$]{
\includegraphics[width=0.51\textwidth]{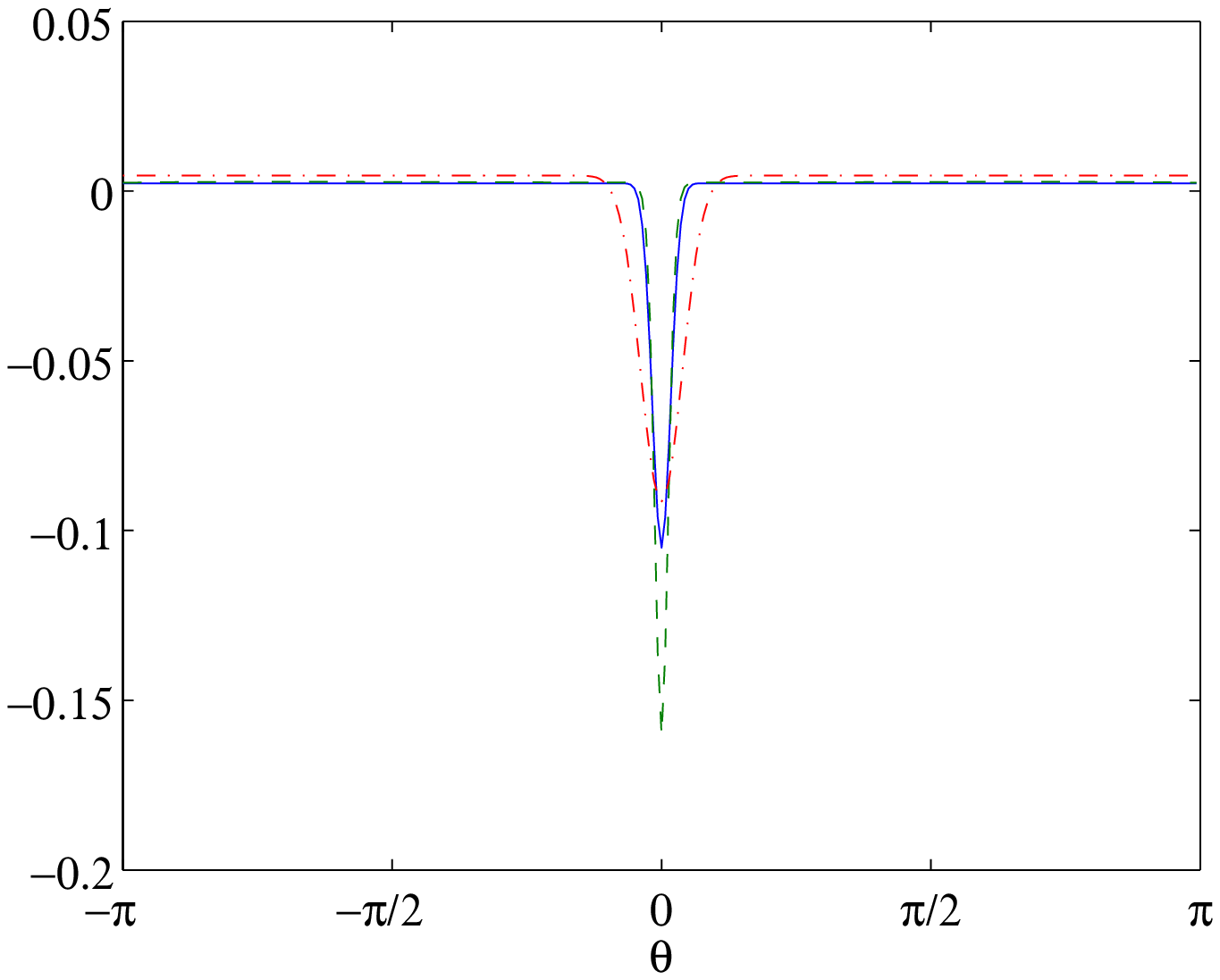}\label{fig:easyvrRe5}}
\subfigure[$-\Dpartial{g}{r}(r_0,\theta)$]{
\includegraphics[width=0.51\textwidth]{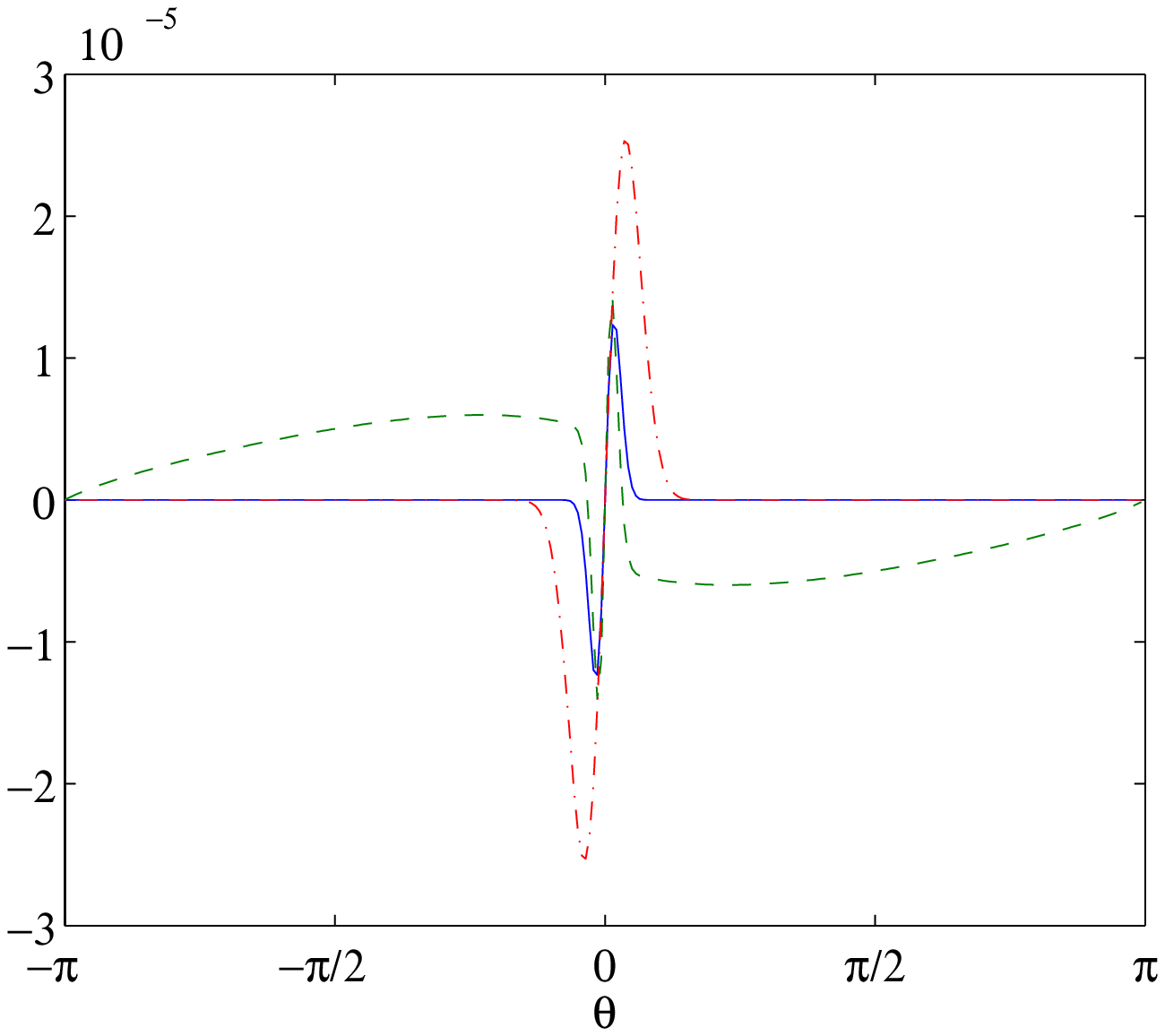}\label{fig:easyvthetaRe5}}}
\caption{Comparison between the approach of B\"{o}nisch et
  al.~\cite{bhw05} (expression \eqref{eq:Wittverstream}, dot-dashed
  line), the present approach (expression \eqref{eq:g2}, solid line)
  and the data obtained from a finite-difference solution of
  Navier-Stokes system \eqref{eq:streamvorticity} without ansatz
  \eqref{eq:psi'} (dashed lines) for the function $g$ and its
  different derivatives appearing in system
  \eqref{eq:streamvorticitybootstrap}.  The Reynolds number is $Re=10$
  and radial distance $r_0=277$.}
\label{fig:g}
\end{figure}

In the present study we follow a different approach to construction of
function $g$. It relies on a simplification and approximation of the
series solution to Oseen equations
\eqref{eq:oseen_a}--\eqref{eq:oseen_b} derived by Tomotika and Aoi
\cite{Tomotika1950} and revisited in \cite{Gustafsson2012}. The
expression obtained is
\begin{equation}
g(r,\theta) = \frac{F_x}{2} \left[ \frac{\theta}{\pi} - \Erf \left( \sqrt{Re_O r} \sin{\theta} \right) \right],
\label{eq:g2}
\end{equation}
and we refer the reader to \cite{gPhD} for a discussion of all
assumptions and steps required in the derivation. In \eqref{eq:g2}
$Re_O$ denotes the equivalent ``Reynolds number'' characterizing the
Oseen flow governed by \eqref{eq:oseen} which has the same drag $F_x$
as the computed Navier-Stokes flow, cf.~\eqref{eq:PRvel}. The value of
the drag in the Oseen flow can be obtained from the data available in
\cite{Gustafsson2012}. To illustrate the properties of expression
\eqref{eq:g2}, in Figure \ref{fig:g} we compare the behaviour of its
different derivatives appearing in system
\eqref{eq:streamvorticitybootstrap} with formula
\eqref{eq:Wittverstream} and the data obtained from a
finite-difference solution of Navier-Stokes problem
\eqref{eq:streamvorticity} without ansatz \eqref{eq:psi'} at some
large radial distance $r$.

\section{Key Ingredients of the Numerical Approach}
\label{sec:numer}

In this Section we discuss the key ingredients of the numerical
approach developed to solve problem
\eqref{eq:streamvorticitybootstrap}, namely, discretization of the
differential operators, imposition of the boundary conditions,
filtering and, finally, the structure of the resulting algebraic
system together with the technique used to solve it. In addition to
the novel approach described below, a standard second-order
finite-difference method was also implemented for comparison and
validation purposes (e.g., see the data shown in Figure \ref{fig:g}).

In order to achieve high accuracy, in the present problem we have
adopted a spectral approach combining the Fourier-Galerkin
discretization in the azimuthal direction with a collocation method
based on the rational Chebyshev polynomials for the discretization in
the radial direction. Since the Fourier-Galerkin method on a periodic
domain is fairly standard \cite{Canuto2007}, we describe it here only
very briefly. The streamfunction $\psi(r,\theta)$ and the vorticity
$\omega(r,\theta)$ are odd functions of the azimuthal angle $\theta$,
hence they can be approximated with sine series
\begin{equation}\label{eq: fourierdiscrete}
\begin{aligned}
\psi(r,\theta) &\approx \sum_{k=1}^{N_1} \hat{\psi}_k(r) \sin(k \theta), \quad r \in[1, \infty], \ \theta \in[0, 2 \pi],\\
\omega(r,\theta) &\approx \sum_{k=1}^{N_1} \hat{\omega}_k(r) \sin(k \theta),
\end{aligned}
\end{equation}
where $N_1>0$ is the number of terms, whereas $\hat{\psi}_k(r)$ and
$\hat{\omega}_k(r)$, $k = 1, \dots, N_1$, are the corresponding
Fourier coefficients depending on the radial coordinate $r$.
Differentiation with respect to the azimuthal angle $\theta$ is
standard and involves multiplication of the corresponding Fourier
coefficients by the wavenumber. Discretization in the radial direction
is more complicated and is described in detail below.

\subsection{Discretization of Functions Defined on a Semi-Infinite Interval Using Rational Chebyshev Functions}
\label{sec:rational}

\begin{figure}
	\centering
\includegraphics[width=0.8\textwidth]{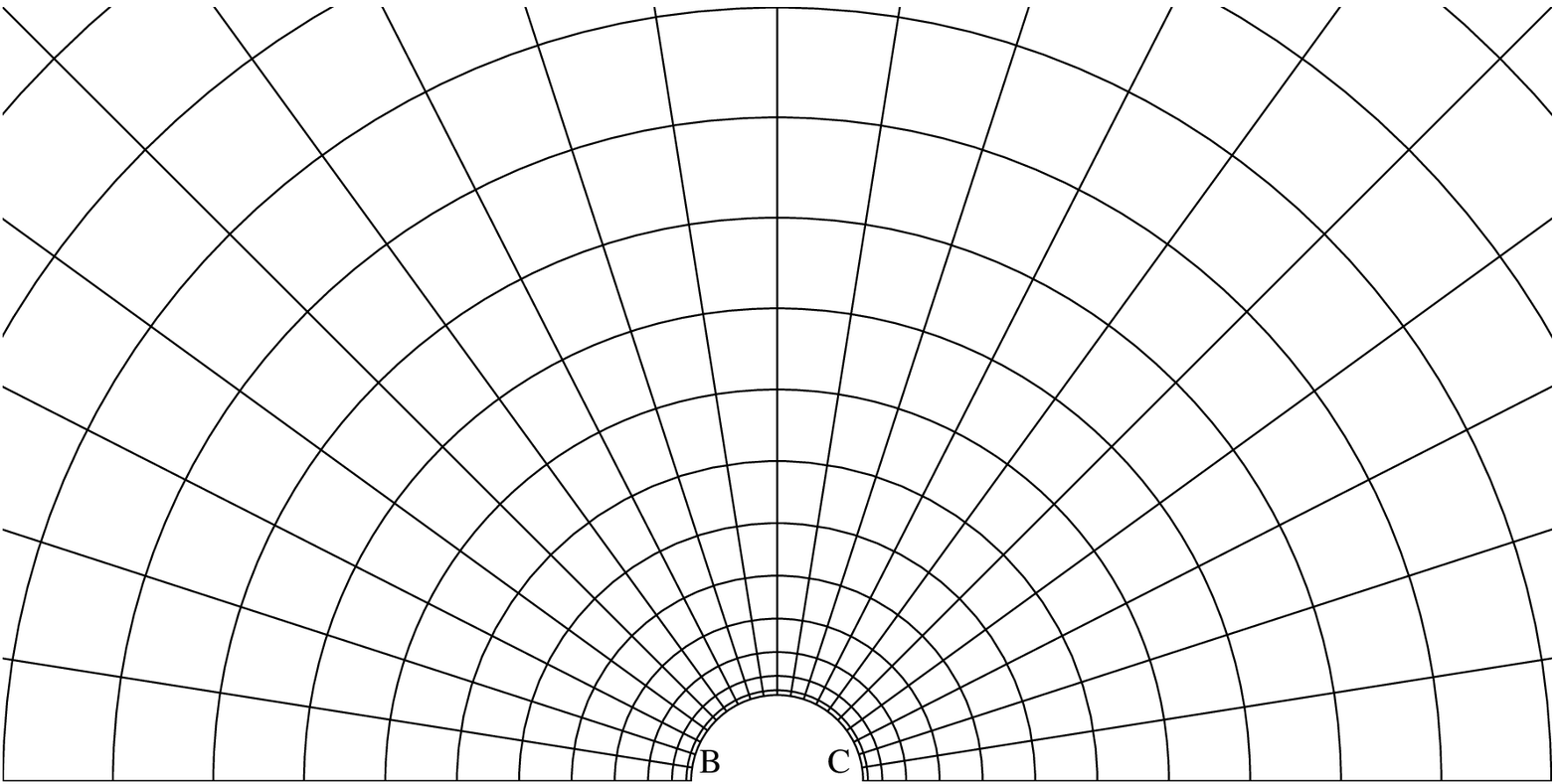}
\includegraphics[width=0.8\textwidth]{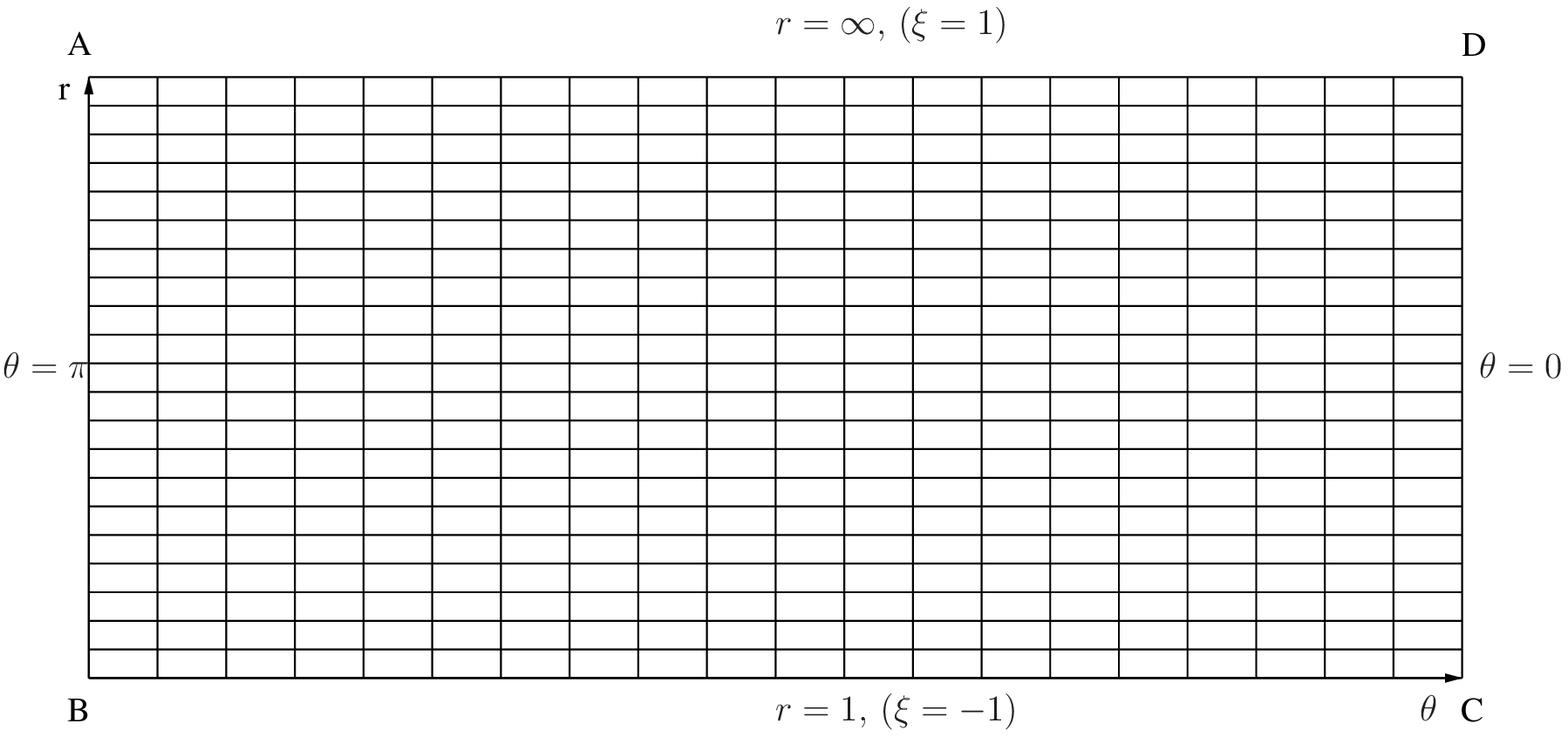}
\caption{Mapping from the exterior $\Omega$ of the cylinder into the
  computational domain $[-1,1] \times [0,2\pi]_{\textrm{per}}$ based
  on transformation \eqref{eq:map}. For clarity, a uniform
  discretization of variable $\xi$ is used.}
\label{fig:cmapping}
\end{figure}
Computational methods based on rational Chebyshev functions,
introduced by Grosch \& Orszag \cite{go77} and Boyd \cite{b82},
belong to a broader family of coordinate transformation approaches.  In
contrast to other families of basis functions defined on unbounded
intervals, such as the ``sinc'' functions or the Hermite polynomials,
the rational Chebyshev functions are characterized by a slow
(algebraic) decay when the argument becomes large (functions from the
former families vanish exponentially fast \cite{Boyd2000}). This
property is particularly important in the present problem, where we
need to obtain solutions with a prescribed slow decay at infinity.
Using the rational Chebyshev functions can be seen as a two-step
process. First, we map the semi-unbounded region $\R := [1,\infty]$ to
$\I := [-1,1]$. Using variables $r \in \R$ and $\xi \in \I$ this
mapping can be realized, for example, through the following algebraic
transformation \cite{Boyd2000}
\begin{equation}\label{eq:map}
r(\xi) = \frac{L(1+\xi)+1-\xi}{1-\xi}\quad \Longleftrightarrow \quad \xi(r) = \frac{r-L-1}{r+L-1},
\end{equation}
where $L>0$ is a parameter. Mapping \eqref{eq:map} is schematically
illustrated in Figure \ref{fig:cmapping}, where we show the
transformation of the (unbounded) flow domain $\Omega = \RR \backslash
A$ to the computational domain $[-1,1] \times [0,2\pi]_{\textrm{per}}$
(the subscript ``per'' implies that the interval is periodic). The
second step is to expand the unknown functions (i.e., the Fourier
coefficients $\hat{\psi}_k(r)$ and $\hat{\omega}_k(r)$, $k = 1, \dots,
N_1$, of the streamfunction and vorticity) in terms of the rational
Chebyshev functions $TL_n(r)$, $n=1,\dots$. They are defined by
composing Chebyshev polynomials $T_n(\xi)$ with mapping \eqref{eq:map}
\begin{equation}
TL_n(r(\xi)) := T_n(\xi), \quad r \in{[1, \infty]}, \quad \xi \in{[-1,1]}.
\label{eq:TL}
\end{equation}
Finally, the expansions are truncated by retaining $N_2$ terms and
collocated at the Gauss-Lobatto points in $[-1,1]$ \cite{Peyret2002}
\begin{equation}\label{eq:collocation}
\xi_i = \cos\left(\frac{i\pi}{N_2-1}\right), \quad i=0,\ldots,N_2-1,
\end{equation}
resulting in 
\begin{equation}\label{eq:psiw}
\begin{aligned}
\psi(r,\theta) &\approx \sum_{k=1}^{N_1} \sum_{n=0}^{N_2-1} \hat{\psi}_{k,n} \sin(k \theta), \\
\omega(r,\theta) &\approx \sum_{k=1}^{N_1} \sum_{n=0}^{N_2-1} \hat{\omega}_{k,n} \sin(k \theta),
\end{aligned}
\end{equation}
where $\hat{\psi}_{k,n} := \hat{\psi}_{k}(\xi_n)$ and
$\hat{\omega}_{k,n} := \hat{\omega}_{k}(\xi_n)$ are the values
of the sine series expansions coefficients at the collocation points
\eqref{eq:collocation}. In regard to transformation \eqref{eq:map},
according to \cite{Boyd2000}, the parameter $L$ should be chosen equal
to the characteristic scale of variation of the solution. Half of the
collocation points will be between 1 and $1+L$, and the other half
will be between $1+L$ and infinity. The first collocation point
$\xi_0$ is mapped to $r=\infty$. As concerns the transformation of the
derivatives, mapping \eqref{eq:map} has the following property
\begin{subequations}\label{eq:derivativer}
\begin{align}
\prfrac{} &= \frac{Q(\xi)}{2L} \frac{\partial}{\partial \xi},& \xi \in{[-1,1]}, \\
\frac{\partial^2}{\partial r^2} &= \frac{(Q(\xi))^2}{4L^2} \frac{\partial^2}{\partial \xi^2} 
+ \frac{2(\xi-1)Q(\xi)}{4L^2}\frac{\partial}{\partial \xi},
\end{align} 
\end{subequations}
where,
\begin{equation}\label{eq:Q}
Q(\xi) := \xi^2-2\xi+1, \quad \xi \in{[-1,1]},
\end{equation}
whereas the approximation of the derivatives with respect to the
transformed variable $\xi$ follows the rules of the spectral Chebyshev
differentiation \cite{Peyret2002} and is described briefly in Appendix
\ref{sec:chebdiff}. Tests validating this method are presented in
Section \ref{sec:NStest}. We refer the reader to monograph
\cite{Boyd2000} for more details concerning the history, properties
and applications of rational Chebyshev functions.

\subsection{Filtering}
\label{sec:filter}

As was shown in Section \ref{sec:perturbation}, the steady solutions we
are looking for have the property that their streamfunction
$\psi(r,\theta)$ has a jump discontinuity in the coordinate $\theta$
as $r \rightarrow \infty$ (cf.~Figure \ref{fig:g}a). While the idea of
the proposed method is to isolate this singularity in the field
$g(r,\theta)$ which is then subtracted off, cf.~\eqref{eq:psi'}, it
may happen, especially at early iterations of Newton's method (see
below), that the magnitude of $g$ (which is proportional to the drag
force $F_x$) is not chosen correctly and a finite jump discontinuity
may also be present in the field $\psi'$. Resolving this singularity
while using a spectral representation of the solution in an unbounded
domain poses a number of challenges.  More precisely, due to this
discontinuity, for sufficiently large wavenumbers $k$ the Fourier
coefficients in \eqref{eq:psiw} may become unbounded with $r$. Given
its global character, cf.~Appendix \ref{sec:chebdiff}, Chebyshev
differentiation may become unstable leading to small-scale
oscillations. For Chebyshev polynomials the error will be concentrated
at $x = -1,1$ and will increase with $N_1$ \cite{Breuer1992}, whereas
for rational Chebyshev functions the error will be largest at $x=-1$,
i.e., at the surface of the cylinder. To ensure smoothness, it is
therefore necessary to apply filtering. One possibility is to apply
the filter to the solution \cite{Canuto2007}.  We will instead use the
{\it derivative filtering} suggested in \cite{Majda1978,Kreiss1979}
which regularizes the discrete derivative operators ${\bf \bar{D}}^1$
and ${\bf \bar{D}}^2$. Such regularization is typically performed in
the spectral space and the first-order differentiation matrix ${\bf
  \bar{D}}^1$, cf.~\eqref{eq:D1}, is replaced with
\begin{equation}\label{eq:filteredderivative}
{\bf T} \cdot {\bf\bar{D}}^1 \cdot \diag{(\tilde{\sigma})} \cdot {\bf T}^{-1},
\end{equation}
where ${\bf T}$ is the matrix representing the transformation from the
physical to the spectral-Chebyshev space, ${\bf T}^{-1}$ its inverse
and
\begin{equation}
\tilde{\sigma} = \left[ \sigma_0, \sigma_1, \ldots, \sigma_{N_1-1} \right]
\label{eq:sigma}
\end{equation} 
is the vector of the discrete filter values obtained as
\begin{equation}
\sigma_k = \sigma \left(\frac{k}{N_1-1}\right), \quad k=0,\ldots,N_1-1.
\label{eq:sigmak}
\end{equation}
The two most popular filters are defined as
\cite{Canuto2007,Vandeven1991}
\begin{subequations}
\label{eq:filters}
\begin{align}
\sigma(\theta) &= e^{-\alpha \theta^p}, \label{eq:expfilter}\\
\sigma(\theta) &= 1 - \frac{(2p-1)!}{(p-1)!^2}\int_0^{\theta /\pi} t^{p-1} (1-t)^{p-1}\,dt, \label{eq:intfilter}
\end{align}
\end{subequations}
where $\alpha$ and $p$ are parameters. Expression \eqref{eq:expfilter}
is referred to as an ``exponential filter''. The coefficient $p$
should be even and positive, whereas $\alpha$ should be chosen so that
$\sigma(1) \simeq 0$. Note that, if $\sigma_k = 1$ for
$k=0,\ldots,N_1-1$, then expression \eqref{eq:filteredderivative} is
the same as ${\bf \bar{D}}^{(1)}$ down to round-off errors. We also
note that filters \eqref{eq:filters} can increase the condition number
of the differentiation matrix by a few orders of magnitude resulting
in a slower convergence of Newton's method. In Section
\ref{sec:results} we comment on the choice of the filter parameters
and the effect they have on the computed solutions.

\subsection{Algebraic System Corresponding to Discretization of Problem \eqref{eq:streamvorticitybootstrap}}
\label{sec:system}

We now describe how the discretization approaches introduced above can
be used to derive an algebraic form of system
\eqref{eq:streamvorticitybootstrap}. To simplify notation, we
introduce the following vectors
\begin{subequations}
\begin{align}
\boldsymbol{\hat{\psi}} = [\boldsymbol{\hat{\psi}}_{1}, \ldots ,\boldsymbol{\hat{\psi}}_{N_1}]^T, \qquad
\boldsymbol{\hat{\psi}}_k &= [\hat{\psi}_{k,0}, \ldots ,\hat{\psi}_{k,N_2-1}]^T, \ k=1,\dots,N_1, \\
\boldsymbol{\hat{\omega}} = [\boldsymbol{\hat{\omega}}_{1}, \ldots ,\boldsymbol{\hat{\omega}}_{N_1}]^T, \qquad 
\boldsymbol{\hat{\omega}}_k &= [\hat{\omega}_{k,0}, \ldots ,\hat{\omega}_{k,N_2-1}]^T, \\
{\bf \bar{D}}^{(1)}_i &= [\bar{d}^{(1)}_{i,0}, \ldots, \bar{d}^{(1)}_{i,N_2-1}], \label{eq: D1galerkin}\\
{\bf \bar{D}}^{(2)}_i &= [\bar{d}^{(2)}_{i,0}, \ldots, \bar{d}^{(2)}_{i,N_2-1}]. \label{eq: D2galerkin}
\end{align}
\end{subequations}
The vectors $\boldsymbol{\hat{\psi}}_k$ and
$\boldsymbol{\hat{\omega}}_k$ represent, respectively, the values of
the $k$-th terms in sine series \eqref{eq:psiw} at the different
collocation points. The vectors ${\bf \bar{D}}^{(1)}_i$ and ${\bf
  \bar{D}}^{(2)}_i$ allow us to approximate, respectively, the first
and second derivative with respect to $\xi$ at the collocation points
$\xi_i$, cf.~Appendix \ref{sec:chebdiff}. We also need to provide a compatible
representation of the ``skeleton'' function $g(r,\theta)$ appearing in
system \eqref{eq:streamvorticitybootstrap}. We add that this
function is chosen to capture the leading-order behavior
of the streamfunction field far from the obstacle and
as such may provide a rather poor representation of the flow field
close to the obstacle. To mitigate the effect this can have on the
numerical solution, we introduce a ``mask'' $H \; : \; \RR \rightarrow
\RR^+$ which will smoothly damp the skeleton function away from the wake region, so that
\begin{equation}
\tg(r,\theta) := H(r) \, g(r,\theta)
\label{eq:mask}
\end{equation}
will replace $g$ in the discrete version of
problem \eqref{eq:streamvorticitybootstrap}. The mask function $H(r)$
ought to satisfy the following conditions
\begin{subequations}
\label{eq:mask2}
\begin{align}
\lim_{r\rightarrow \infty} H(r) &= 1, \label{eq:maska} \\
H(1) = H'(1) = H''(1) = H^{(3)}(1) &= 0, \label{eq:maskb}
\end{align}
\end{subequations}
where \eqref{eq:maskb} are chosen to ensure that the
mask does not affect the drag force, cf.~\eqref{eq:dragVeysey}. The
following mask will be used
\begin{equation}\label{eq:filterH}
H(r) = \frac{1+\Erf\left[\kappa (r-R_{1/2})\right]}{2},
\end{equation}
where $\kappa$ is a parameter characterizing the localization of the
mask and $R_{1/2}$ is the distance at which the mask reduces the
amplitude of the background term by half, i.e., $H(R_{1/2}) = 1/2$. We
note that function \eqref{eq:filterH} satisfies conditions
\eqref{eq:maskb} approximately with improving accuracy as $\kappa$
increases. Moreover, as $\kappa \rightarrow \infty$, we have
\begin{equation}
H(r) = \mathcal{H}(r-R_{1/2}). 
\end{equation}
Since the skeleton function enters into system
\eqref{eq:streamvorticitybootstrap} only through terms involving
derivatives, we will need the following expansions
\begin{subequations}\label{eq:filterbootstrap}
\begin{align}
H(r) \frac{1}{r} \frac{\partial \tilde{g}}{\partial \theta} &= F_x \sum_{k=1}^{N_g} a_k(r) \cos(k \theta) \label{eq:filterbootstrapone},\\
-H(r)\frac{1}{r} \frac{\partial \tilde{g}}{\partial r} -H'(r)\frac{\tilde{g}}{r} &= F_x \sum_{k=1}^{N_g} b_k(r) \sin(k\theta), \label{eq:filterbootstraptwo}
\end{align}
\begin{equation}
H(r)\Delta \tilde{g}+\left[ H^{(2)}(r) +\frac{H'(r)}{r} \right] \tilde{g}+2H'(r) \frac{\partial \tilde{g}}{\partial r} 
 = F_x \sum_{k=1}^{N_g}c_k(r) \sin(k \theta), \label{eq:filterbootstrapthree}
\end{equation}
\end{subequations}
where $\{a_k\}_{k=1}^{N_g}$, $\{b_k\}_{k=1}^{N_g}$, and
$\{c_k\}_{k=1}^{N_g}$, $N_g > 0$, are expansion coefficients
determined in a standard way for any choice of the function $g$,
cf.~Sections \ref{sec:Wittver} and \ref{sec:asymptoticoseen}. We
emphasize that, regardless of the specific choice of the skeleton $g$,
expansions \eqref{eq:filterbootstrap} depend linearly on the drag
force $F_x$ which in turn depends on the entire solution to the
problem, cf.~\eqref{eq:dragFornberg}--\eqref{eq:dragVeysey}.
Substituting representation \eqref{eq:psiw} together with ansatz
\eqref{eq:mask} and expansions \eqref{eq:filterbootstrap} into system
\eqref{eq:streamvorticitybootstrap} and following standard steps
(collocation in the transformed radial direction $\xi$ and Galerkin
approach in the azimuthal direction $\theta$; see \cite{gPhD} for all
details), we obtain a nonlinear algebraic problem of the form
\begin{equation}\label{eq:W}
\boldsymbol{W} \left( \left[ \begin{array}{c}
\boldsymbol{\hat{\psi}} \\
\boldsymbol{\hat{\omega}}
\end{array} \right] \right) +{\bf A} \left[ \begin{array}{c}
 \boldsymbol{\hat{\psi}} \\
 \boldsymbol{\hat{\omega}}
\end{array} \right] +F_x \, {\bf B} \left[ \begin{array}{c}
 \boldsymbol{\hat{\psi}} \\
 \boldsymbol{\hat{\omega}}
\end{array} \right]= {\bf 0}.
\end{equation}
It consists of $N_1 \times N_2$ equations in the same number of
unknowns. The operator $\bW$ is a nonlinear function of
$\boldsymbol{\hat{\psi}}$ and $\boldsymbol{\hat{\omega}}$ which it is
convenient to split as
\begin{equation}\label{eq:W12}
\boldsymbol{W} \left( \left[ \begin{array}{c}
\boldsymbol{\hat{\psi}} \\
\boldsymbol{\hat{\omega}}
\end{array} \right] \right) = \bW^{(1)} \left( \left[ \begin{array}{c}
\boldsymbol{\hat{\psi}} \\
\boldsymbol{\hat{\omega}}
\end{array} \right] \right)+ \bW^{(2)} \left( \left[ \begin{array}{c}
\boldsymbol{\hat{\psi}} \\
\boldsymbol{\hat{\omega}}
\end{array} \right] \right),
\end{equation}
where
\begin{multline}
\left[ \boldsymbol{W}^{(1)} \left( \left[ \begin{array}{c}
\boldsymbol{\hat{\psi}} \\
\boldsymbol{\hat{\omega}}
\end{array} \right] \right) \right]_{j+N_2 (l+k-1)} = \\
\frac{Q(\xi_j)}{4Lr(\xi_j)} \left( k \hat{\psi}_{k,j} {\bf D}^{(1)}_j \cdot \boldsymbol{\hat{\omega}_l} -  l \hat{\omega}_{l,j} {\bf D}^{(1)}_j \cdot \boldsymbol{\hat{\psi}}_{k} \right),
\end{multline}
with the subscript on the left-hand side (LHS) enumerating the rows.
The index $j$ corresponds to the collocation points $\xi_j$, whereas
$l$ and $k$ are the wavenumbers in sine series expansions
\eqref{eq:psiw}.  Terms in the sine series expansions corresponding to
$k+l \ge N_1$, which originate from the quadratic nonlinearity, are
not resolved and are truncated. As regards the second term in equation
\eqref{eq:W12}, we have
\begin{multline}
\left[ \boldsymbol{W}^{(2)} \left( \left[ \begin{array}{c}
\boldsymbol{\hat{\psi}} \\
\boldsymbol{\hat{\omega}}
\end{array} \right] \right) \right]_{j+N_2 (\sgn(l-k)(l-k)-1)} = \\
 \sgn (l-k) \frac{Q(\xi_j)}{4Lr(\xi_j)} \left( k \hat{\psi}_{k,j}  {\bf D}^{(1)}_j \cdot \boldsymbol{\hat{\omega}}_l   + l \hat{\omega}_{l,j} {\bf D}^{(1)}_j \cdot \boldsymbol{\hat{\psi}}_{k} \right),
\end{multline}
where $\sgn(x)$ is the signum (sign) function. The linear part of
system \eqref{eq:W} consists of operator $\bA$, corresponding to the
dissipative term in the momentum equation
\eqref{eq:streamvorticitybootstrapa} and the kinematic relation
\eqref{eq:streamvorticitybootstrapb} between the streamfunction and
vorticity, which has the following form
\begin{multline}\label{eq:Agalerkin}
\left[ {\bf A} \left[ \begin{array}{c}
 \boldsymbol{\hat{\psi}} \\
 \boldsymbol{\hat{\omega}}
\end{array} \right] \right]_{j+N_2(l-1)} = \\
\frac{\mathcal{H}[l-2]}{2}\left[\frac{Q(\xi_j)}{2L} {\bf D}^{(1)}_j \cdot \boldsymbol{\hat{\omega}}_{l-1} -\frac{l-1}{r(\xi_j)} \hat{\omega}_{l-1,j}\right] \\
+\frac{1-\mathcal{H}[l-N_1]}{2} \left[ \frac{Q(\xi_j)}{2L} {\bf D}^{(1)}_j \cdot \boldsymbol{\hat{\omega}}_{l+1} +\frac{l+1}{r(\xi_j)} \hat{\omega}_{l+1,j} \right] \\
-\frac{2}{Re} \frac{Q(\xi_j)}{2L}\left[ \frac{Q(\xi_j)}{2L} {\bf D}^{(2)}_j + \frac{2(\xi_j-1)}{2L} {\bf D}^{(1)}_j + \frac{1}{r(\xi_j)} {\bf D}^{(1)}_j\right] \cdot \boldsymbol{\hat{\omega}}_{l} -\frac{(l-1)^2}{r(\xi_j)^2} \hat{\omega}_{l,j},
\end{multline}
where $\mathcal{H}[n]$ is the discrete Heaviside step function,
\begin{equation}
\mathcal{H}[n] = \begin{cases}0, & n<0, \\
1, & n \geq 0.
\end{cases}
\end{equation}
The operator $\bB$ corresponds to the terms in system
\eqref{eq:streamvorticitybootstrap} involving the skeleton function
$\tg$ through expansions \eqref{eq:filterbootstrap}. It can be
represented as follows
\begin{equation}\label{eq:B}
{\bf B} \left[ \begin{array}{c}
 \boldsymbol{\hat{\psi}} \\
 \boldsymbol{\hat{\omega}}
\end{array} \right] = {\bf B}^{(1)} \left[ \begin{array}{c}
 \boldsymbol{\hat{\psi}} \\
 \boldsymbol{\hat{\omega}}
\end{array} \right] +{\bf B}^{(2)} \left[ \begin{array}{c}
 \boldsymbol{\hat{\psi}} \\
 \boldsymbol{\hat{\omega}}
\end{array} \right],
\end{equation}
where,
\begin{equation}
\left[ {\bf B}^{(1)} \left[ \begin{array}{c}
 \boldsymbol{\hat{\psi}} \\
 \boldsymbol{\hat{\omega}}
\end{array} \right] \right]_{j+N_2 (k+l-1)}= \frac{1}{2} \left[  a_k(r(\xi_j)) \frac{Q(\xi_j)}{2L} {\bf D}^{(1)}_j \cdot \boldsymbol{\hat{\omega}_l} +l b_k(r(\xi_j)) \hat{\omega}_{l,j} \right], \label{eq:Bone}
\end{equation}
\begin{multline}
\left[ {\bf B}^{(2)} \left[ \begin{array}{c}
 \boldsymbol{\hat{\psi}} \\
 \boldsymbol{\hat{\omega}}
\end{array} \right] \right]_{j+N_2 (\sgn(l-k) (l-k)-1)}= \\
\frac{\sgn (l-k)}{2} \left[  a_k(r(\xi_j)) \frac{Q(\xi_j)}{2L} {\bf D}^{(1)}_j \cdot \boldsymbol{\hat{\omega}_l} -l b_k(r(\xi_j)) \hat{\omega}_{l,j} \right]. \label{eq:Btwo}
\end{multline}
Finally, the boundary conditions in
\eqref{eq:streamvorticitybootstrap} are incorporated in algebraic
system \eqref{eq:W} by replacing the rows corresponding to $j=0$ and
$j=N_2-1$ with the discrete versions of relations
\eqref{eq:streamvorticitybootstrapc}--\eqref{eq:streamvorticitybootstrapf}.
The drag force $F_x$ in equation \eqref{eq:W} is expressed using the
discrete form of relation \eqref{eq:dragFornberg}
\begin{equation}
F_x = \frac{\pi}{Re} \left[ {\bf D}^{(1)}_0 \cdot \boldsymbol{\hat{\omega}}_{1}  -\hat{\omega}_{1,0} \right]
\label{eq:Fx}
\end{equation}
and is updated after a fixed number of iterations.

Denoting $\bX = \left[ \boldsymbol{\hat{\psi}} \ \
  \boldsymbol{\hat{\omega}} \right]^T$, system \eqref{eq:W} can be
rewritten as $\bF(\bX) = \0$, where $\F \; : \; \RR^{N_1N_2}
\rightarrow \RR^{N_1N_2}$ is a nonlinear function. It is solved using
Newton's method
\begin{equation}\label{eq:iter}
{\bf X}^{n+1} = {\bf X}^n- \left[ \bnabla\F({\bf X}^n)\right]^{-1} {\bf F}({\bf X}^n), \quad n=0,1,\ldots 
\end{equation}
in which $\bX^n$ denotes an approximation of the solution obtained at
the $n$-th iteration and the Jacobian $\bnabla\bF({\bf X}^n)$ is
evaluated analytically using trigonometric identities (see \cite{gPhD}
for details). The linear problem required to determine Newton's
direction $- \left[ \bnabla\bF({\bf X}^n)\right]^{-1} {\bf F}({\bf
  X}^n)$ is solved using the LU decomposition (algorithm {\tt dgesv}
from BLAS \cite{LAPACK}). In order to ensure the convergence of
iterations \eqref{eq:iter} away from the solution, a globalization
strategy was used based on adjusting the relative step length
$\alpha$, so that it would satisfy the minimization condition
\cite{Kelley2003}
\begin{equation}
\min_{\alpha \in{[0,1]}}  \big\|{\bf F} ({\bf X}^n- \alpha \, \left[ \bnabla\bF({\bf X}^n)\right]^{-1}  {\bf F}({\bf X}^n))\big\|_2.
\end{equation}
This problem is solved using the function {\tt fmin} from {\tt
  netlib}, modified so as to find the minimum along the direction $-
\left[ \bnabla\bF({\bf X}^n)\right]^{-1} {\bf F}({\bf X}^n)$ from the
point ${\bf X}^n$. The algorithm uses the ``golden section'' search
and parabolic interpolation \cite{Forsythe1976}. A number of different
initial guesses $\bX^0$ were used to initialize Newton's iterations,
including the potential flows, Oseen flows and solutions already
obtained at slightly different Reynolds numbers. Newton's iterations
\eqref{eq:iter} were declared converged when the norm of the residual
$\| {\bf F}({\bf X}^n) \|_2$ was reduced below $10^{-9}$.

\section{Results}
\label{sec:results}

In this Section we first present a systematic validation of the
proposed approach on simplified model problems which nonetheless
highlight the issues relevant to our main task. Then, we show
results for the steady Navier-Stokes flows computed for a range of
Reynolds numbers.

\subsection{Validation Tests of the Proposed Numerical Method}
\label{sec:NStest}

\begin{figure}
\mbox{
\subfigure[]{\includegraphics[width=0.48\textwidth]{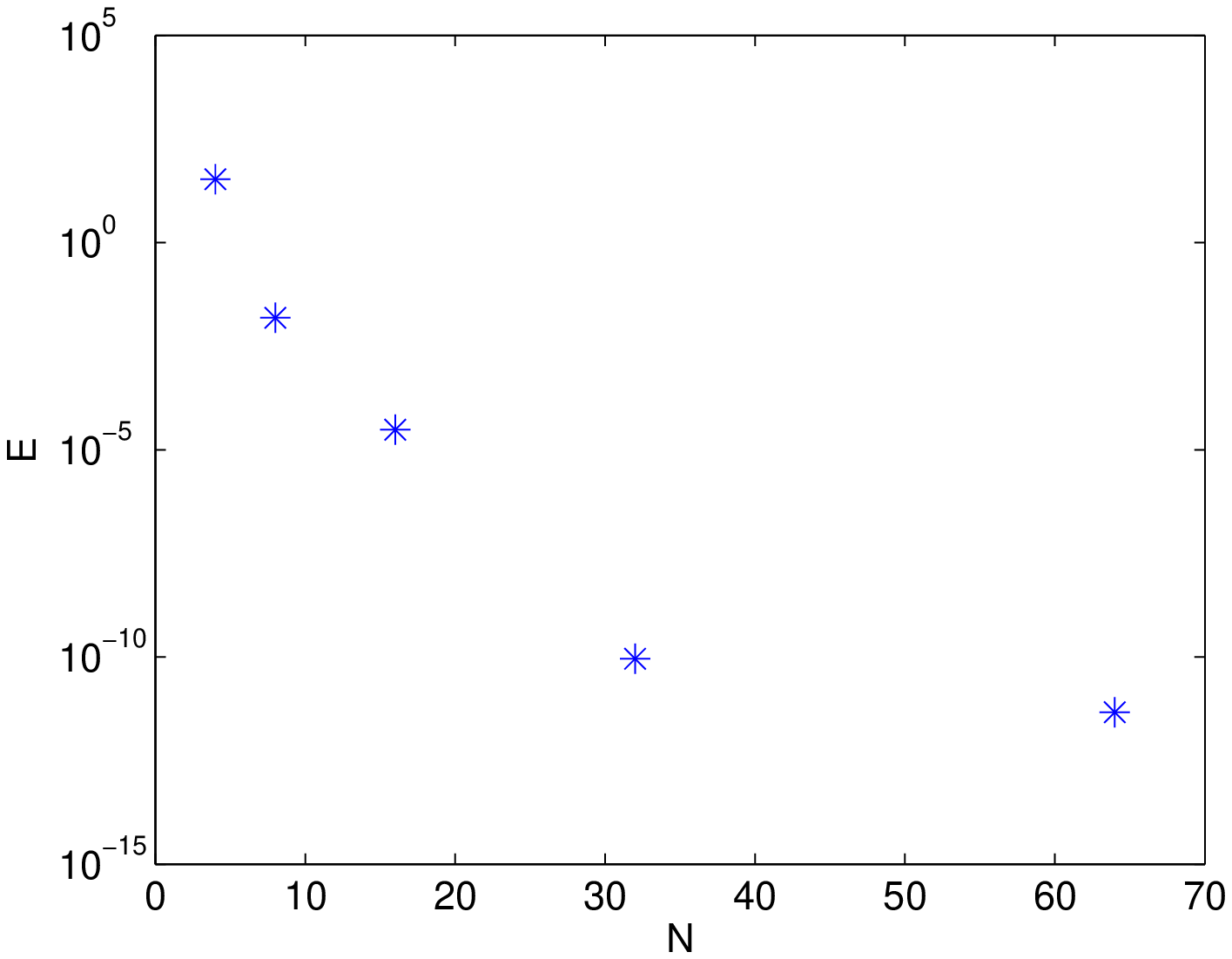}}
\quad
\subfigure[]{\includegraphics[width=0.48\textwidth]{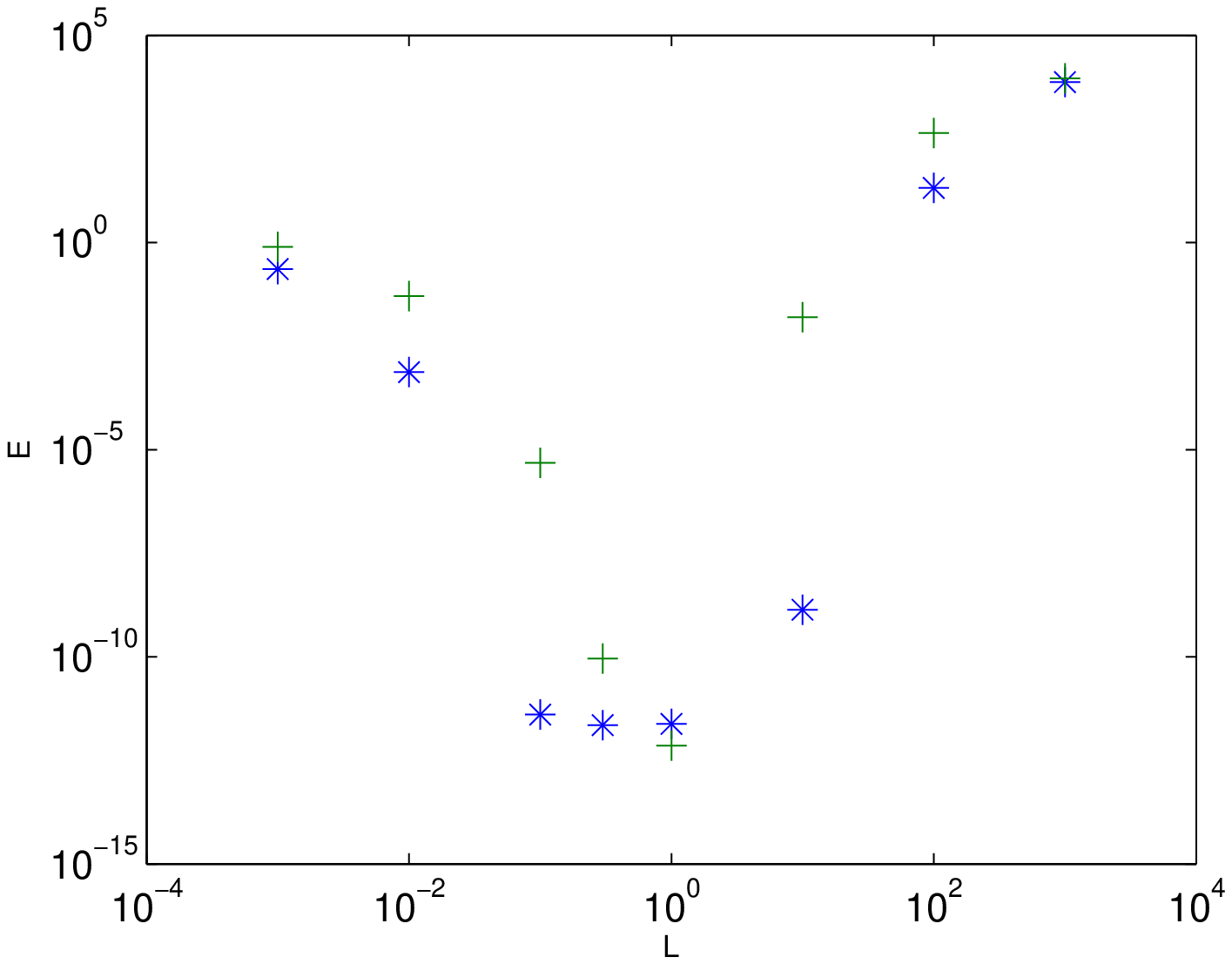}}}
\caption{Error \eqref{eq:Errf} in the solution of model problem
  \eqref{eq:lap} as a function of (a) resolution $N = N_1 = N_2$ with
  a fixed length-scale $L = 0.3$ and (b) length-scale $L$ for two
  fixed resolutions $(+)$ $N=32$ and $(*)$ $N=64$.}
\label{fig:LapErr}
\end{figure}

To demonstrate the consistency of the tools developed in Section
\ref{sec:rational}, we begin by considering Laplace's equation defined
on the same unbounded domain $\Omega$ as Navier-Stokes system
\eqref{eq:streamvorticitybootstrap}, namely,
\begin{subequations}\label{eq:lap}
\begin{align}
  \nabla^2 f &= 0, &\text{in} \ \Omega,\\
f & = \sum_{k=1}^8 \sin(k \theta),&\text{on} \ \partial A,\\
\frac{\partial f}{\partial r} & = -\sum_{k=1}^8 k \sin(k \theta),&\text{on} \ \partial A,
\end{align}
\end{subequations}
whose exact solution is
\begin{equation}
\label{eq:lapsolex}
f^{\textrm{ex}}(r, \theta) = \sum_{k=1}^8 \frac{\sin(k \theta)}{r^k}.
\end{equation}
We remark that, rather than being a boundary-value problem, system
\eqref{eq:lap} is in fact a Cauchy problem for the Laplace equation.
It is known to be ill-posed and therefore represents a more stringent
test for our approach. In addition, such formulation allows us to
validate the simultaneous imposition of two boundary conditions on
the cylinder boundary,
cf.~\eqref{eq:streamvorticitybootstrapc}--\eqref{eq:streamvorticitybootstrapd}.
We solve problem \eqref{eq:lap} assuming
\begin{equation}
\label{eq:lapsol}
f(r,\theta) = \sum_{k=1}^{N_1} {f}_k(r) \sin(k \theta), \quad r \in{[1, \infty)}, \quad \theta \in{[0, 2 \pi)}
\end{equation}
and using the method described in Section \ref{sec:rational}. The error is
defined as
\begin{equation}
E := \max_{k \in{[1, N_1]}, \ i \in{[0,N_2-1]} }| \hat{f}_{k}(r(\xi_i)) - \hat{f}^{\textrm{ex}}_k(r(\xi_i))|,
\label{eq:Errf}
\end{equation}
where $\hat{f}^{\textrm{ex}}_k(r)$ are the Fourier coefficients of
exact solution \eqref{eq:lapsolex}, and Figure \ref{fig:LapErr}a
confirms that the expected spectral accuracy is achieved when
resolution $N := N_1 = N_2$ increases.

An important numerical parameter in the approach developed in Section
\ref{sec:rational} is the length-scale $L$ parameterizing mapping
\eqref{eq:map}. Sensitivity of the accuracy of the results to this
parameter is examined in Figure \ref{fig:LapErr}b which demonstrates
that the errors are smallest for intermediate values of $L$, of order
$\O(1)$, and increase when $L$ is both very small and very large.
While these findings are not quite unexpected \cite{Boyd2000}, data in
Figure \ref{fig:LapErr}b provides quantitative information about the
values of $L$ one should use.

\begin{table}
\centering
\begin{tabular}{*6{c}}
\hline\noalign{\smallskip}
Case \# & $\psi^{\textrm{test}}$            &  $\omega^{\textrm{test}}$  & $E_R$ \\
\hline
1 & 0 & $\frac{\sin{2\theta}}{r^2}$ & $2 \cdot 10^{-11}$  \\
2 & $\frac{\sin{\theta}}{r}$ & $\frac{\sin{2\theta}}{r^2}$ & $7 \cdot 10^{-10}$  \\
3 & $-\frac{\sin{\theta}}{r^2}$ & $\frac{\sin{2\theta}}{r^2}$ & $1 \cdot 10^{-9}$ \\ 
4 & $\frac{\sin{ 64 \theta}}{r}$ & $\frac{\sin{2\theta}}{r^2}$ & $3 \cdot 10^{-9}$ \\ 
5 & 0 & $\frac{\sin{2\theta}}{r}$ & $6 \cdot 10^{-10}$ \\ 
6 & $\frac{\sin{2\theta}}{r^2}$ & $\frac{\sin{4\theta}}{r^3}$ & $6 \cdot 10^{-10}$ \\ 
7 & $\arctan{(r \sin{\theta})}$ & 0 & $7 \cdot 10^{-3}$ \\ 
8 & $\pi \Erf{(-3 \theta \sqrt{r})}+ \theta$ & 0 & $6 \cdot 10^{-1}$ \\ 
\noalign{\smallskip}\hline
\end{tabular}
%\end{center}
\caption{Summary of the different test cases and the corresponding
  errors \eqref{eq:ErrR} probing the sensitivity of the
  discretization \eqref{eq:W} to the behavior of the test fields at
  infinity. The parameters used are $N = N_1 = N_2 = 64$, $Re = 1$ and
  $L = 1$.}
\label{tab:ErrR}
\end{table} 

\begin{figure}
\mbox{
\subfigure[]{\includegraphics[width=0.48\textwidth]{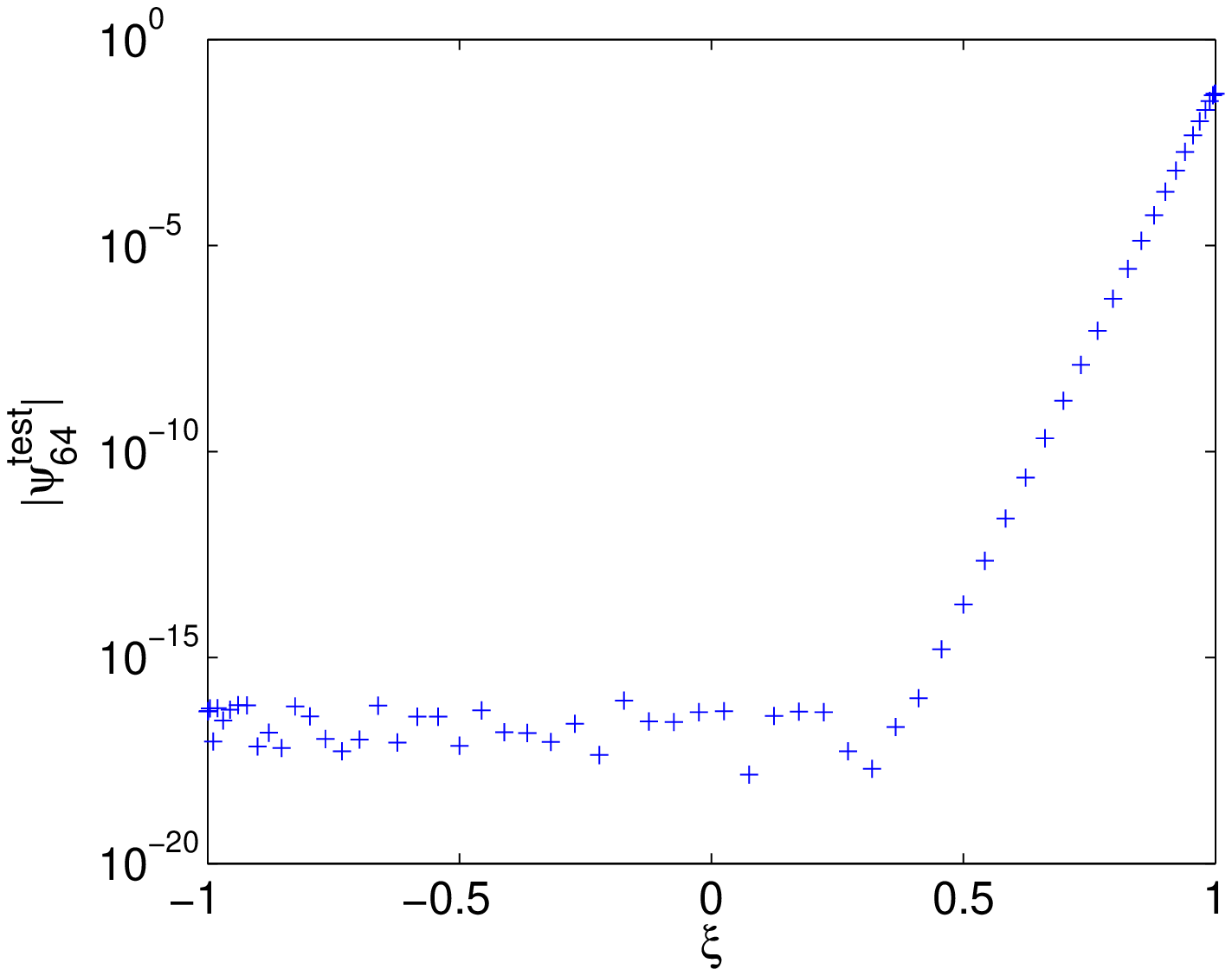}}\quad
\subfigure[]{\includegraphics[width=0.48\textwidth]{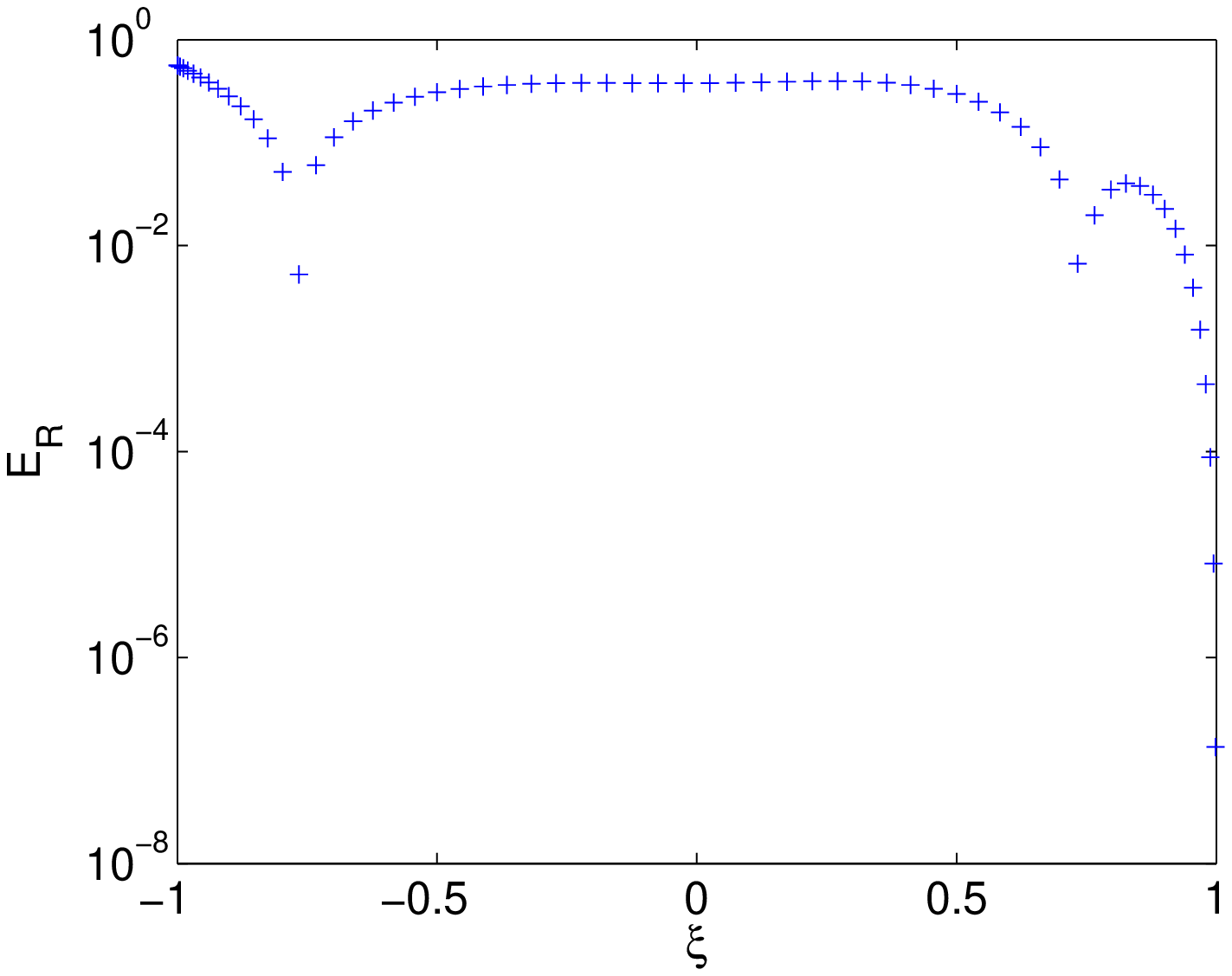}}}
\caption{Dependence of (a) sine series coefficient
  $\hat{\psi}^{\textrm{test}}_{64}$ and (b) error of the second
  derivative $\Dpartialn{\hat{\psi}^{\textrm{test}}_{64}}{\xi}{2}$ with
  respect to the exact values on the rescaled radial distance $\xi$ in
  test case \#8 (cf. Table \ref{tab:ErrR}).}
\label{fig:chebyvsx}
\end{figure}

Next we move on to analyze how the discretization approaches developed
in Section \ref{sec:numer} handle fields characterized by different
behavior at infinity (i.e., slow decay in the radial direction $r$
and/or discontinuity in the azimuthal coordinate $\theta$). This is
motivated by the known properties of the steady Navier-Stokes flows
reviewed in Introduction. This test consists in substituting certain
assumed expressions $\psi^{\textrm{test}}$ and
$\omega^{\textrm{test}}$, respectively, for the streamfunction and
vorticity in discretized system \eqref{eq:W} and comparing the
resulting residuals $\bW( \left[
  \boldsymbol{\hat{\psi}}^{\textrm{test}} \
  \boldsymbol{\hat{\omega}}^{\textrm{test}} \right]^T )$ with the
residual $\bR$ obtained analytically by substituting these expressions
into continuous system \eqref{eq:streamvorticitybootstrap} and then
evaluating it at the collocation points. The error is thus defined as
\begin{equation}
E_R := \Bigg\| \boldsymbol{W} \left( \left[ \begin{array}{c}
\boldsymbol{\hat{\psi}}^{\textrm{test}} \\
\boldsymbol{\hat{\omega}}^{\textrm{test}}
\end{array} \right] \right) - \bR \Bigg\|_{\infty}.
\label{eq:ErrR}
\end{equation}
To focus attention on the effect of the behavior of the different
fields at infinity, numerical resolution and other parameters are
fixed as $N = N_1 = N_2 = 64$, $Re = 1$ and $L = 1$. Information about
different test fields and the corresponding residual errors is
collected in Table \ref{tab:ErrR}, and we refer the reader to
\cite{gPhD} for additional details concerning the analytical forms of
the residual expressions in $\bR$. In Table \ref{tab:ErrR} we observe
that expected accuracy is obtained in all cases except for the last
two. Errors in test case \#7 come from insufficient resolution in the
azimuthal direction far away from the cylinder. Test case \#8 was chosen to capture the wake behaviour expected of
steady 2D Navier-Stokes flows, i.e.,
\begin{equation}
\frac{1}{r} \frac{\partial \psi}{\partial \theta} \sim \frac{1}{\sqrt{r}} \quad \text{when} \ r \rightarrow \infty
\end{equation}
combined with discontinuous dependence on the azimuthal coordinate
$\theta$ (cf.~Sections \ref{sec:Wittver} and
\ref{sec:asymptoticoseen}). It is evident that a straightforward
approach to test cases \#7 and \#8 does not lead to satisfactory
results and, to illustrate the origins of this behavior, in Figure
\ref{fig:chebyvsx}a, we show the dependence of the sine series
coefficient $\hat{\psi}^{\textrm{test}}_{64}$ on the mapped radial
distance $\xi$ in test case \#8.  We note that it is close to zero in
a significant part of the domain extending away from the obstacle
boundary ($\xi=-1$ corresponding to $r=1$, cf.~\eqref{eq:map}), but
becomes quite large when $\xi \rightarrow 1$ corresponding to $r
\rightarrow \infty$.  This behavior stems from the discontinuity of
the streamfunction field in the $\theta$ direction as $r \rightarrow
\infty$ (cf.~Figure \ref{fig:g}a). To show how this behavior affects
the solution process, we evaluate numerically the second derivative
with respect to $r$ of the function shown in Figure
\ref{fig:chebyvsx}a, i.e.,
$\Dpartialn{\hat{\psi}^{\textrm{test}}_{64}}{\xi}{2}$, and in Figure
\ref{fig:chebyvsx}b show the magnitude of the error with respect to
the analytical solution. We see that the errors are quite large,
especially near the obstacle, which given the global nature of
Chebyshev differentiation is a consequence of an unbounded increase of
function $\hat{\psi}^{\textrm{test}}_{64}(\xi)$ as $\xi \rightarrow
1$. Recognizing these issues, the method proposed in this study has
the following two features designed to remedy the underlying problem:
\begin{enumerate}
\item decomposition \eqref{eq:psi'} allows us to effectively contain
  the discontinuity of the streamfunction field $\psi$ in a suitably
  chosen ``skeleton'' function (cf.~Sections \ref{sec:Wittver} and
  \ref{sec:asymptoticoseen}), and
\item spectral filtering of the derivatives described in Section
  \ref{sec:filter}.
\end{enumerate}
Steady Navier-Stokes flows computed employing the above strategy are
discussed in the next Section.

\subsection{Flow Solutions}
\label{sec:flows}

In this Section we present solutions computed for the Reynolds number
spanning the range from 2 to 200. They are obtained with the
resolution $N_1 = 64$, $N_2 = 100$, $N_g=64$ and using $L=1$ as
the parameter of mapping \eqref{eq:map}. Filtering was performed based
on exponential filer \eqref{eq:expfilter} with the parameter $\alpha$
in the range from 12 to 36. Our computational tests indicated that
lower values of $\alpha$ resulted in insufficiently accurate
derivative values which made it difficult to impose Neumann boundary
conditions \eqref{eq:streamvorticitybootstrapd}. On the other hand,
larger values of $\alpha$ resulted in spurious oscillations of the
streamfunction and vorticity fields. The key parameters characterizing
the solutions obtained at different Reynolds numbers are summarized in
Table \ref{tab:flows} where we also list some global diagnostic
quantities typically used to characterize separated wake flows,
namely, the length $L_R$ and half-width $W_R$ of the recirculation
zone, separation angle $\theta_0$ and the drag coefficient $c_D := F_x
/ [(1/2) \rho \, U_{\infty}^2 \, d]$ where $\rho = 1$ (computed in two
different ways). The obtained flow patters are presented in Figure
\ref{fig:flows} in which one can see the isolines of the
streamfunction and vorticity fields together with the corresponding
velocity vector fields. The boundaries of the separated regions are
presented in Figure \ref{fig:recirc}, whereas in Figure
\ref{fig:vorticityonthesurface} one can see the profiles of the
surface vorticity $\omega|_{\partial A}$ as a function of the
azimuthal angle $\theta$. To complete the picture, in Figures
\ref{fig:surfacevorticity}a and \ref{fig:surfacevorticity}b we present
surface plots of the vorticity in the near wake region in the case of
lower ($Re = 20$) and higher ($Re = 200$) Reynolds numbers. These
plots illustrate the evolution of the vorticity field as the Reynolds
number increases, in particular, the emergence of thin shear layers
separating from the obstacle. Finally, in Figure \ref{fig:FDmisc} we
present the dependence of diagnostic quantities $L_R$, $W_R$,
$\theta_0$ and $c_D$ on the Reynolds number and compare them with the
results available in the literature. We note that all the diagnostic
quantities, expect for the length and half-width of the recirculation
region at the highest Reynolds number, exhibit the right trends and
have correct values. The discrepancies appearing at $Re=200$ are
related to insufficient numerical resolution.

\begin{table}
\centering
\begin{tabular}{*{10}{l}}
\hline\noalign{\smallskip}
$Re$ & $N_1$ & $N_2$ & $L$ & Filter           &  $L_R$   & $W_R$ & $\theta_0$ & \multicolumn{2}{c}{$c_D$}\\
     &       &   &   &     $e^{-\alpha t^8}$           &          &      &            & eq. \eqref{eq:dragFornberg}& eq. \eqref{eq:dragVeysey}\\ 
\hline
%\tableheadseprule\noalign{\smallskip}
2 & 64 & 100 & 1 & 12&-      & -      & -      & 6.6772 & 6.6930\\
10 & 64 & 100 & 1 & 19 &1.4264 & 0.7506 & 40.860 & 5.8568 & 5.8412\\
20 & 64 & 100 & 1 & 36 &2.0007 & 0.7722 & 44.629 & 2.2193 & 2.2124\\
100 & 64 & 100 & 1 & 15&13.062 & 1.4820 & 66.832 & 1.1255 & 1.1283\\
200 & 64 & 100 &3 & 15 & 20.006 & 2.4869 & 75.079 & 0.8617 & 0.8629\\
\noalign{\smallskip}\hline
\end{tabular}
%\end{center}
\caption{Numerical parameters used in the computations of steady Navier-Stokes 
flows at different Reynolds numbers and the obtained values of the diagnostic 
quantities (the flow corresponding to $Re=2$ does not exhibit a recirculation zone).
\label{tab:flows}}
\end{table}

\begin{figure}\centering
\mbox{
\subfigure[$Re = 2$]{\plotstream{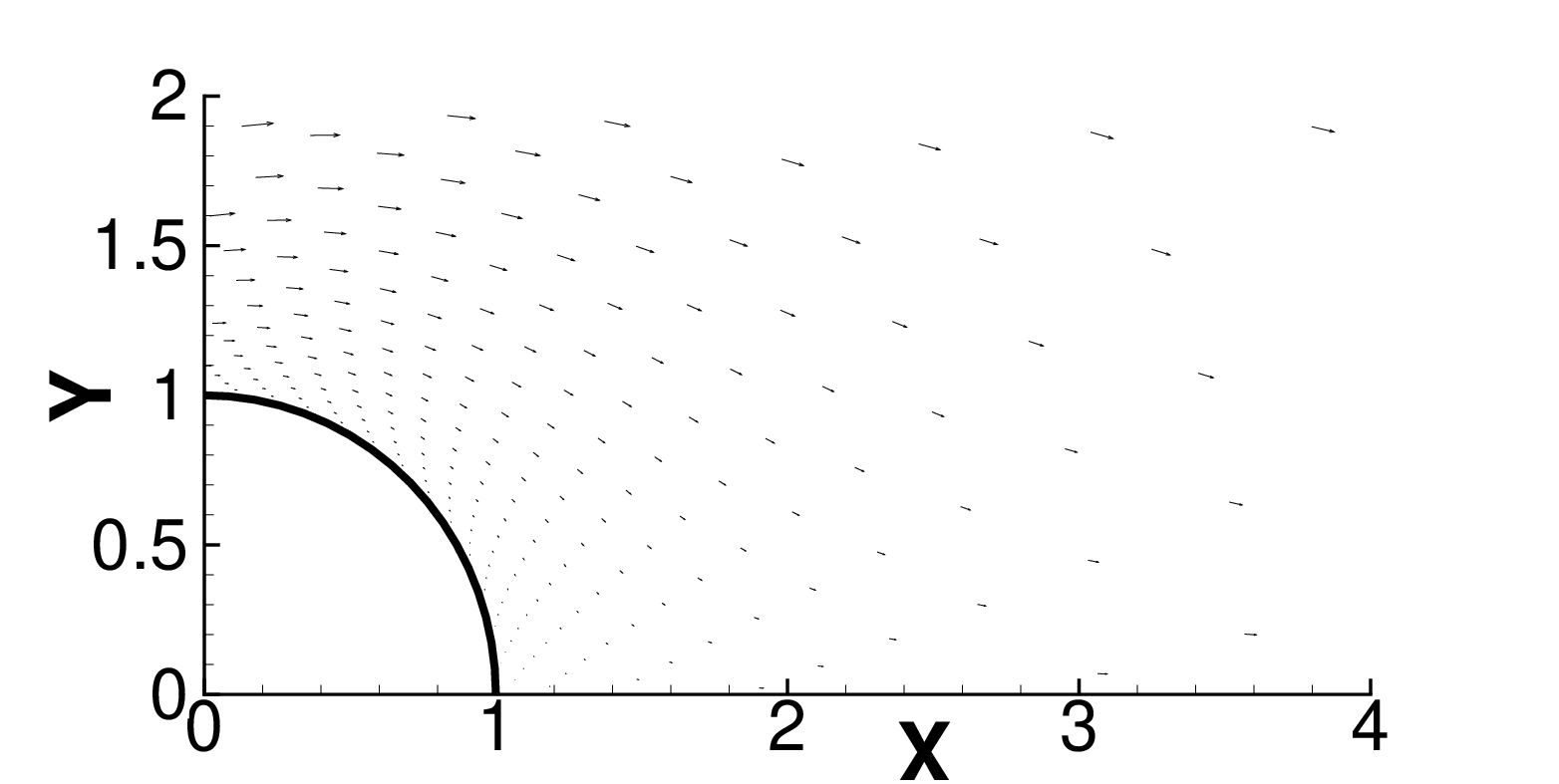}}
\subfigure[$Re = 2$]{\plotvorticity{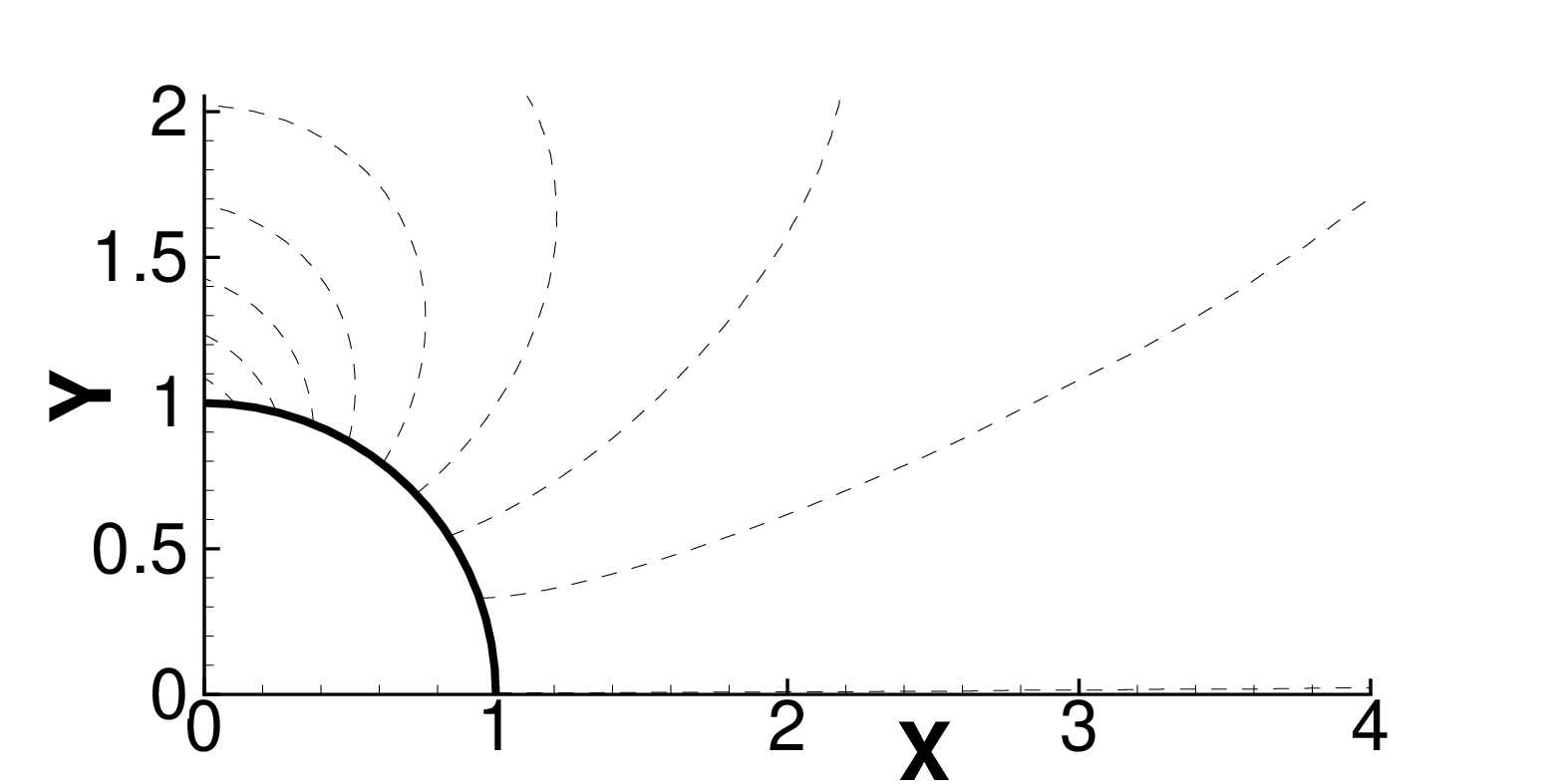}}}
\mbox{\subfigure[$Re = 10$]{\plotstream{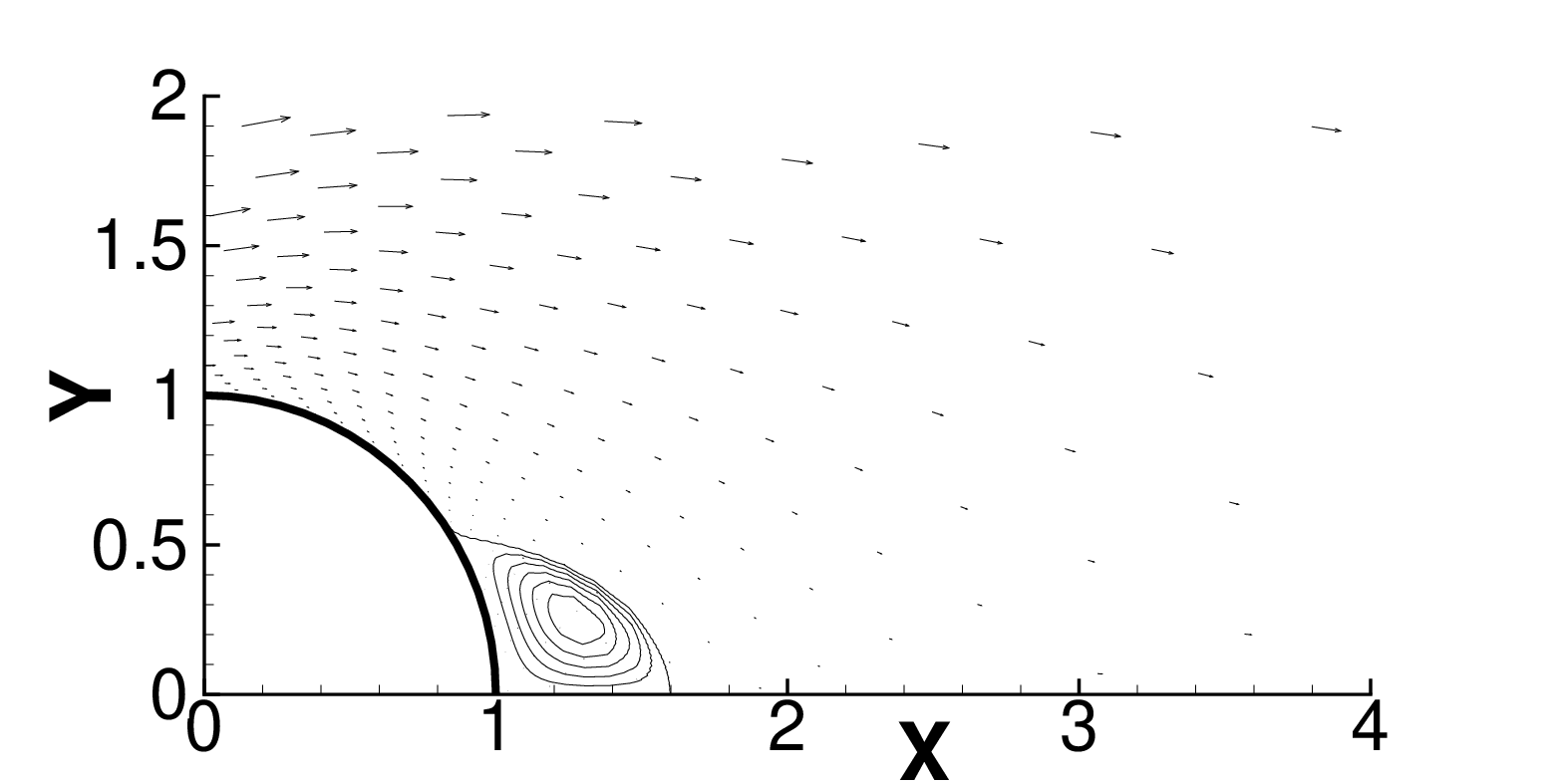}}
\subfigure[$Re = 10$]{\plotvorticity{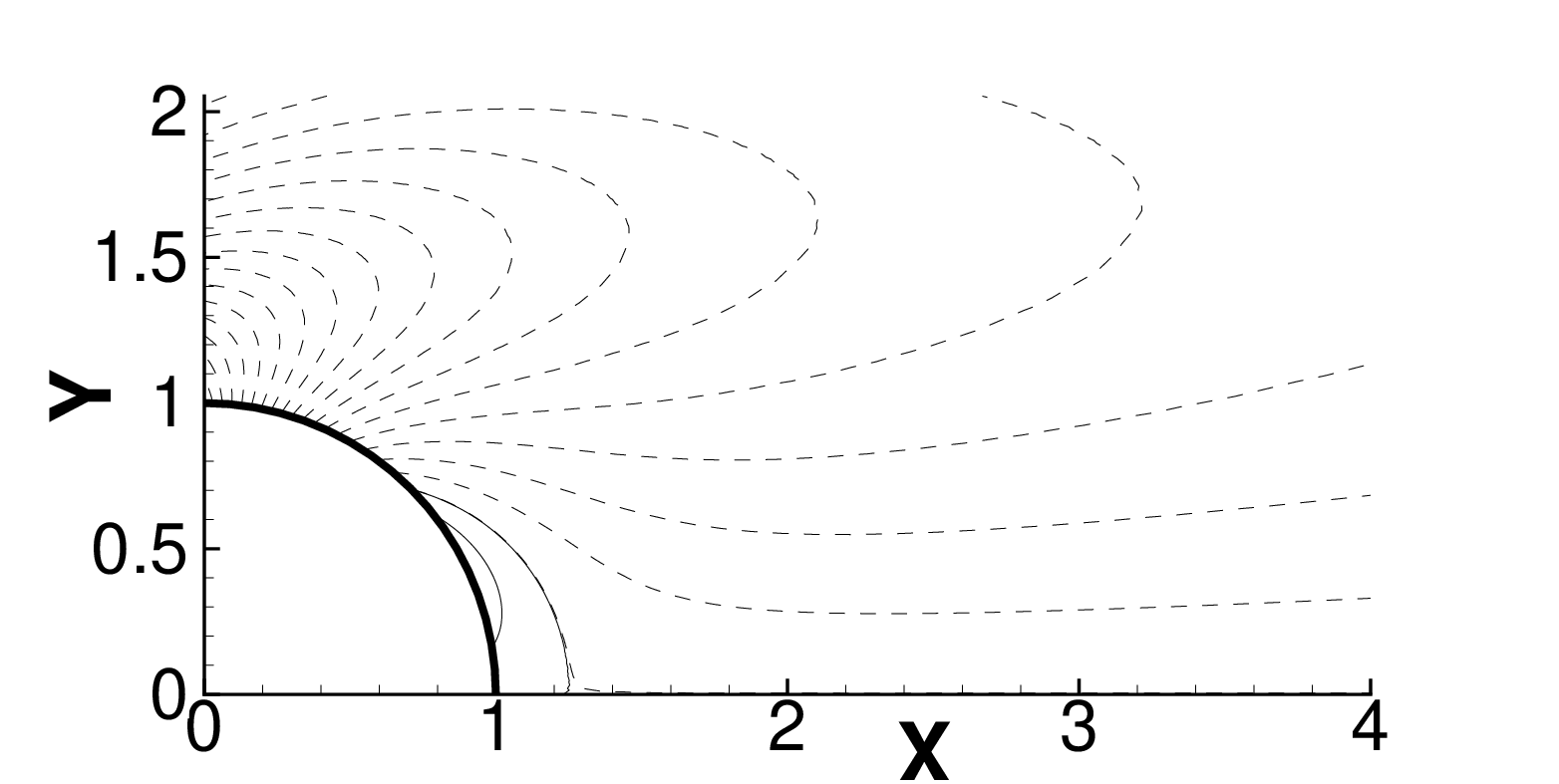}}}
\mbox{\subfigure[$Re = 20$]{\plotstream{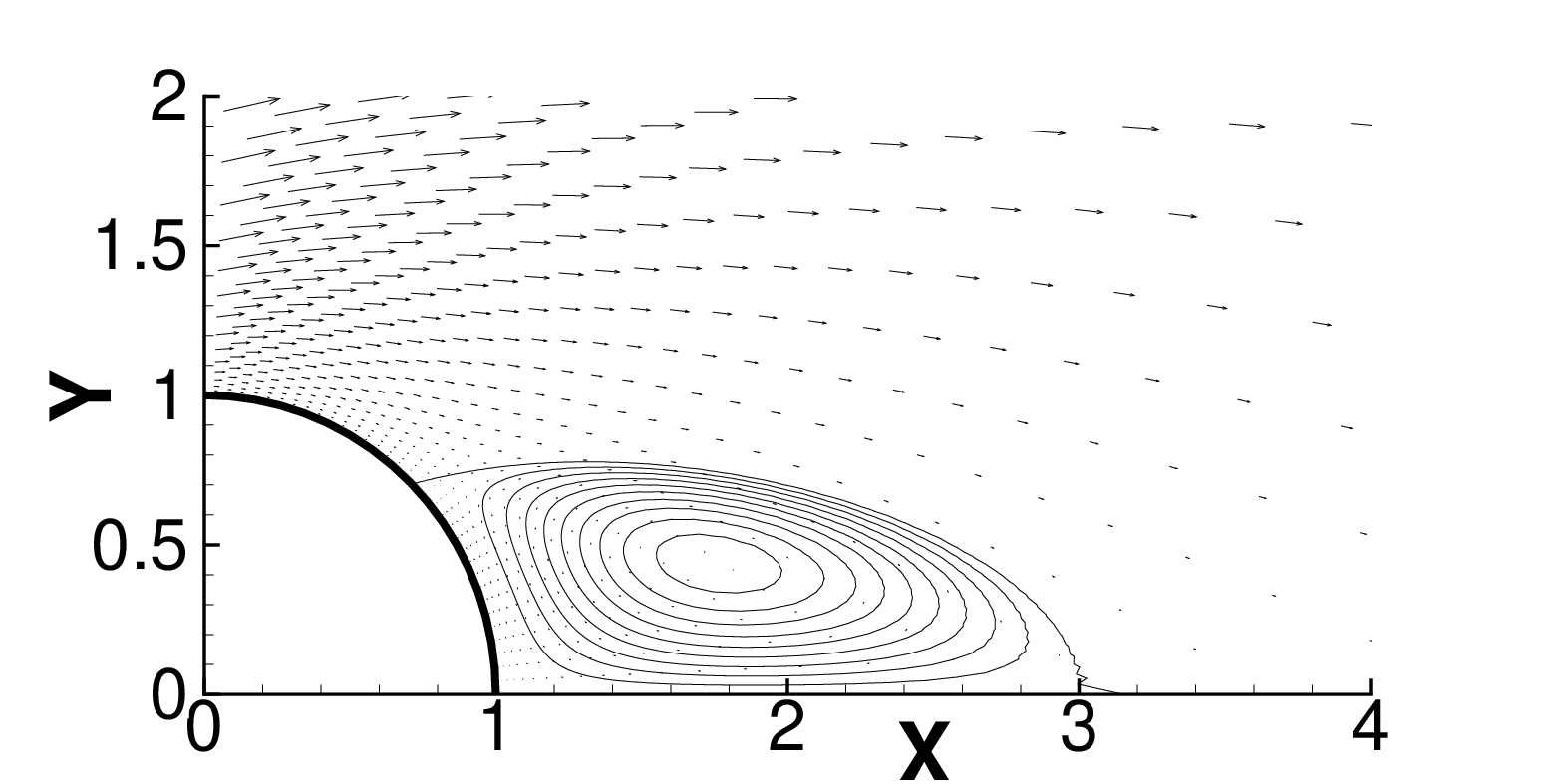}}
\subfigure[$Re = 20$]{\plotvorticity{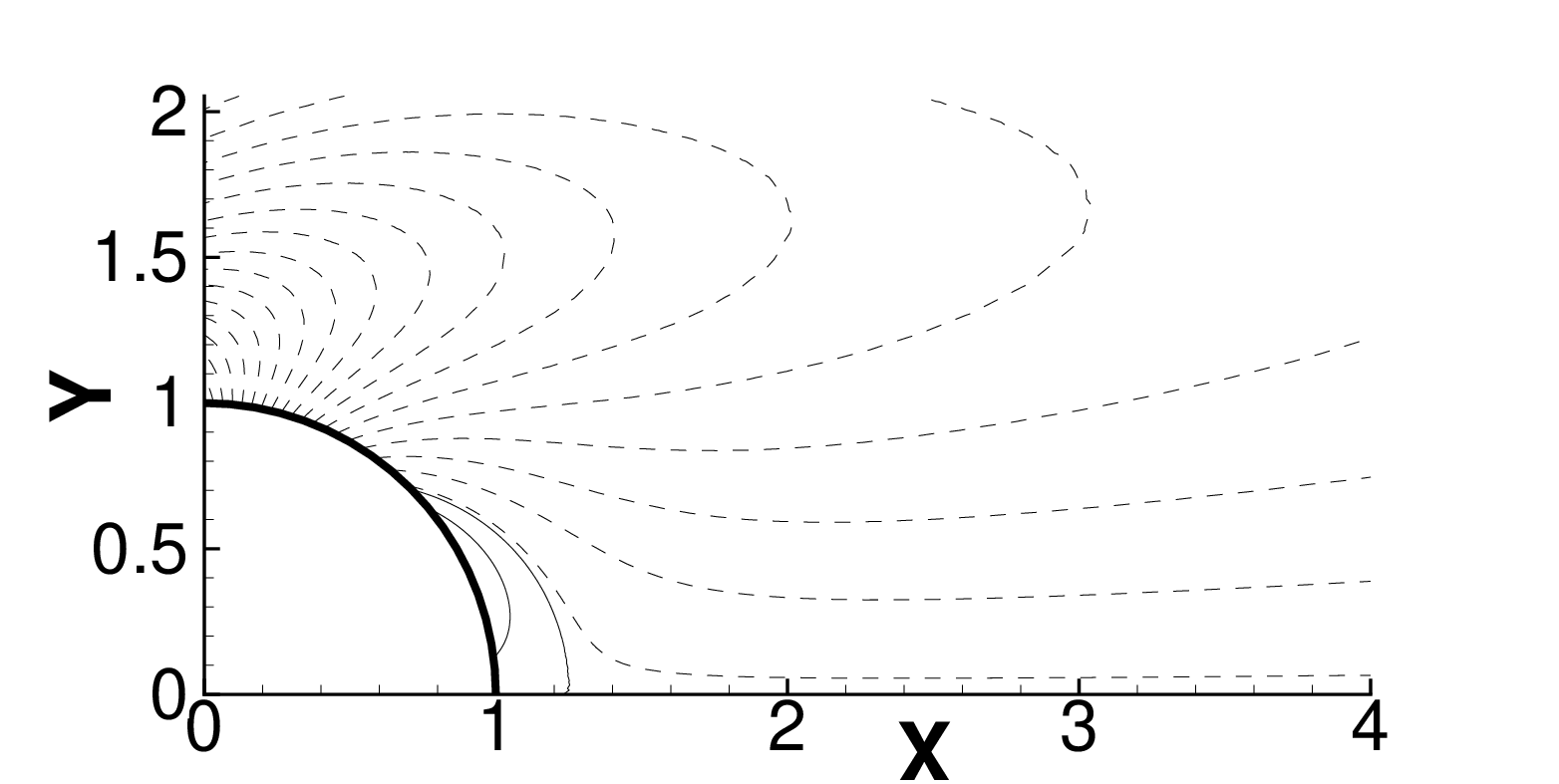}}}
\caption{(Left column) velocity fields and streamline patterns and
  (right column) vorticity fields for the steady Navier--Stokes flows
  obtained for the Reynolds numbers indicated; for clarity, the
  streamlines are shown only in the separated regions; in the
  vorticity plots isocontours corresponding to positive and negative
  vorticity are indicated with solid and dashed lines,
  respectively.\label{fig:flows}}
\end{figure}\addtocounter{figure}{-1}
\begin{figure}\centering
\setcounter{subfigure}{6}
\mbox{\subfigure[$Re = 100$]{\plotstream{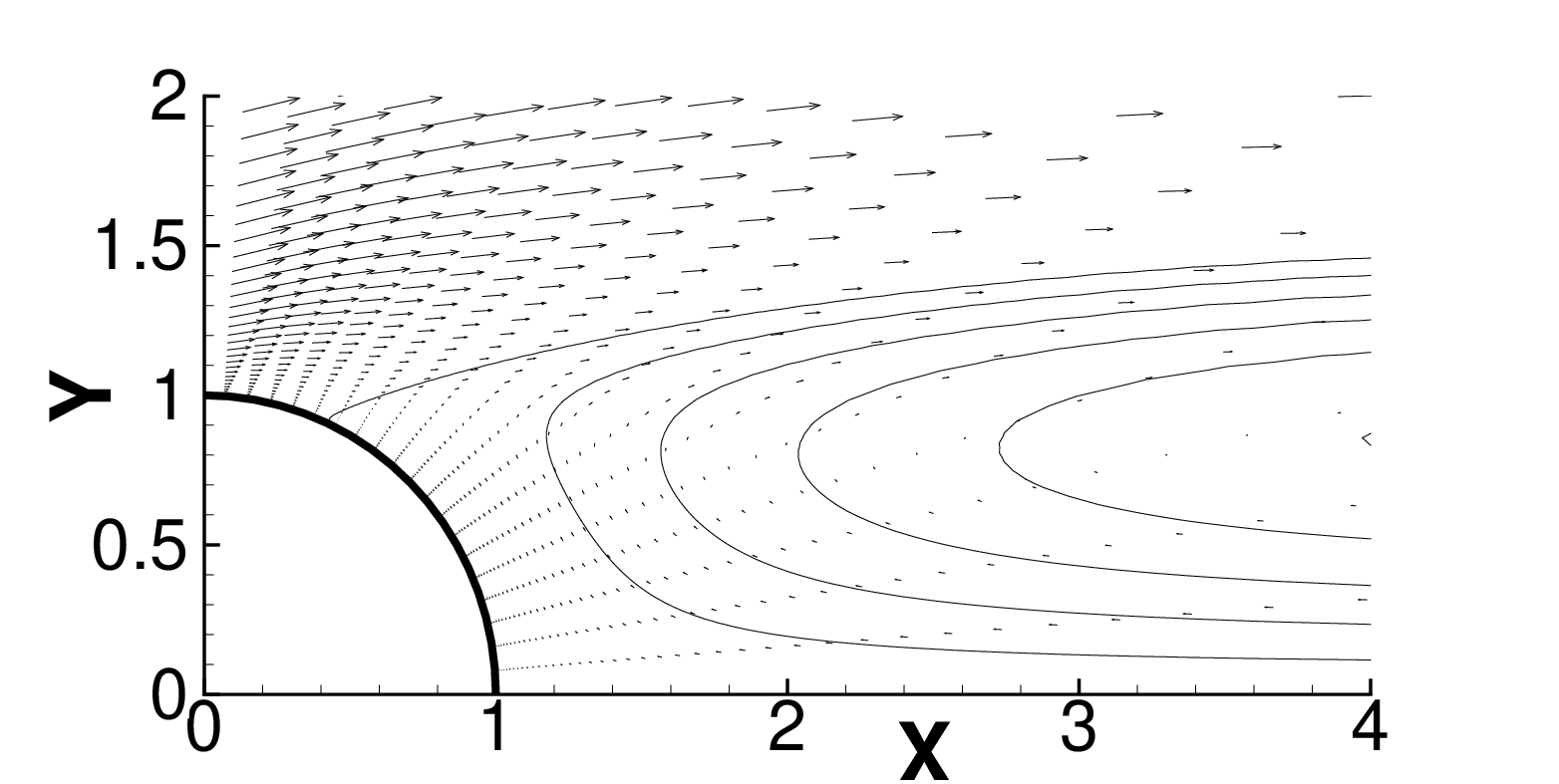}}
\subfigure[$Re = 100$]{\plotvorticity{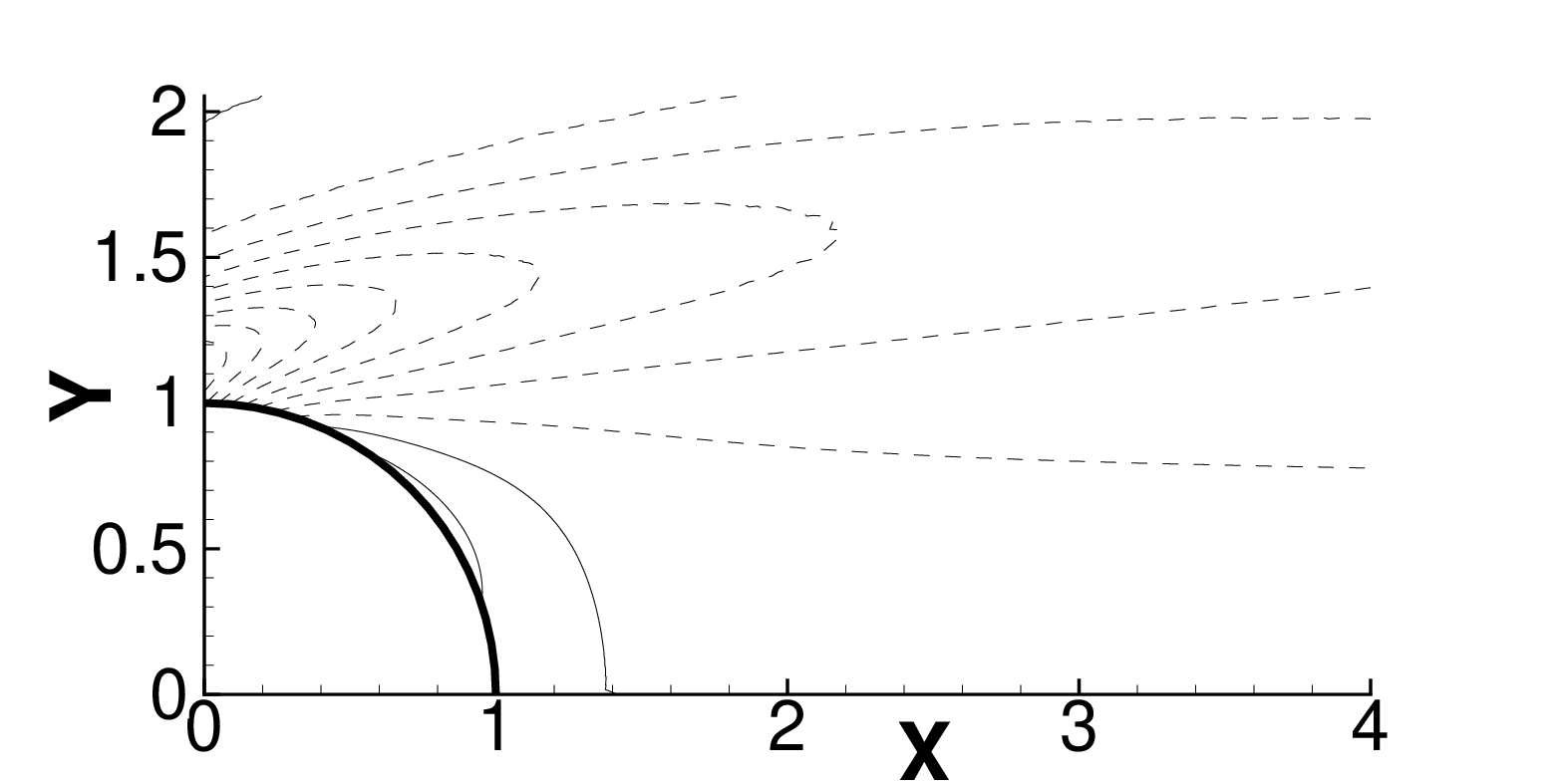}}}
\mbox{\subfigure[$Re = 200$]{\plotstream{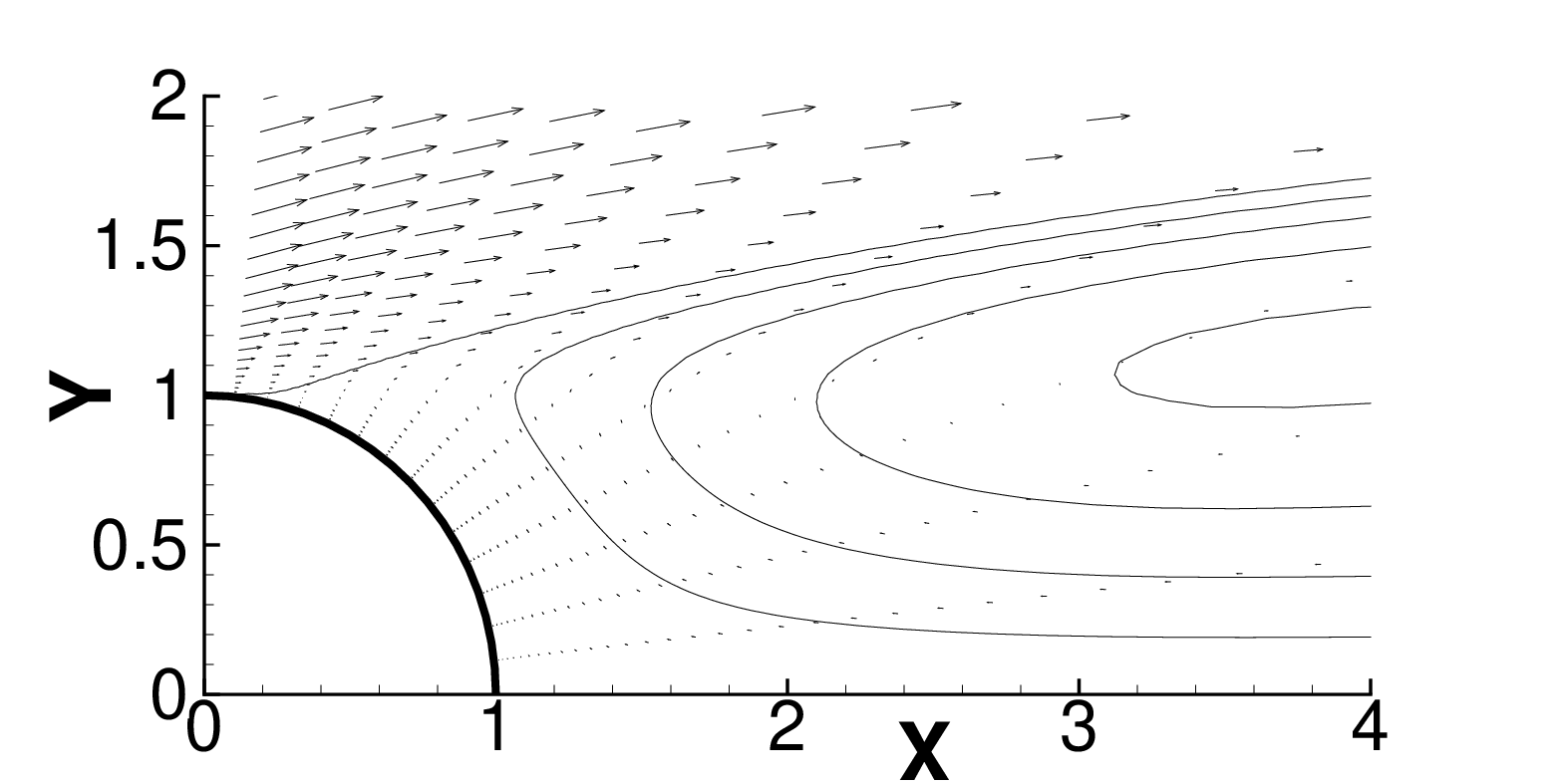}}
\subfigure[$Re = 200$]{\plotvorticity{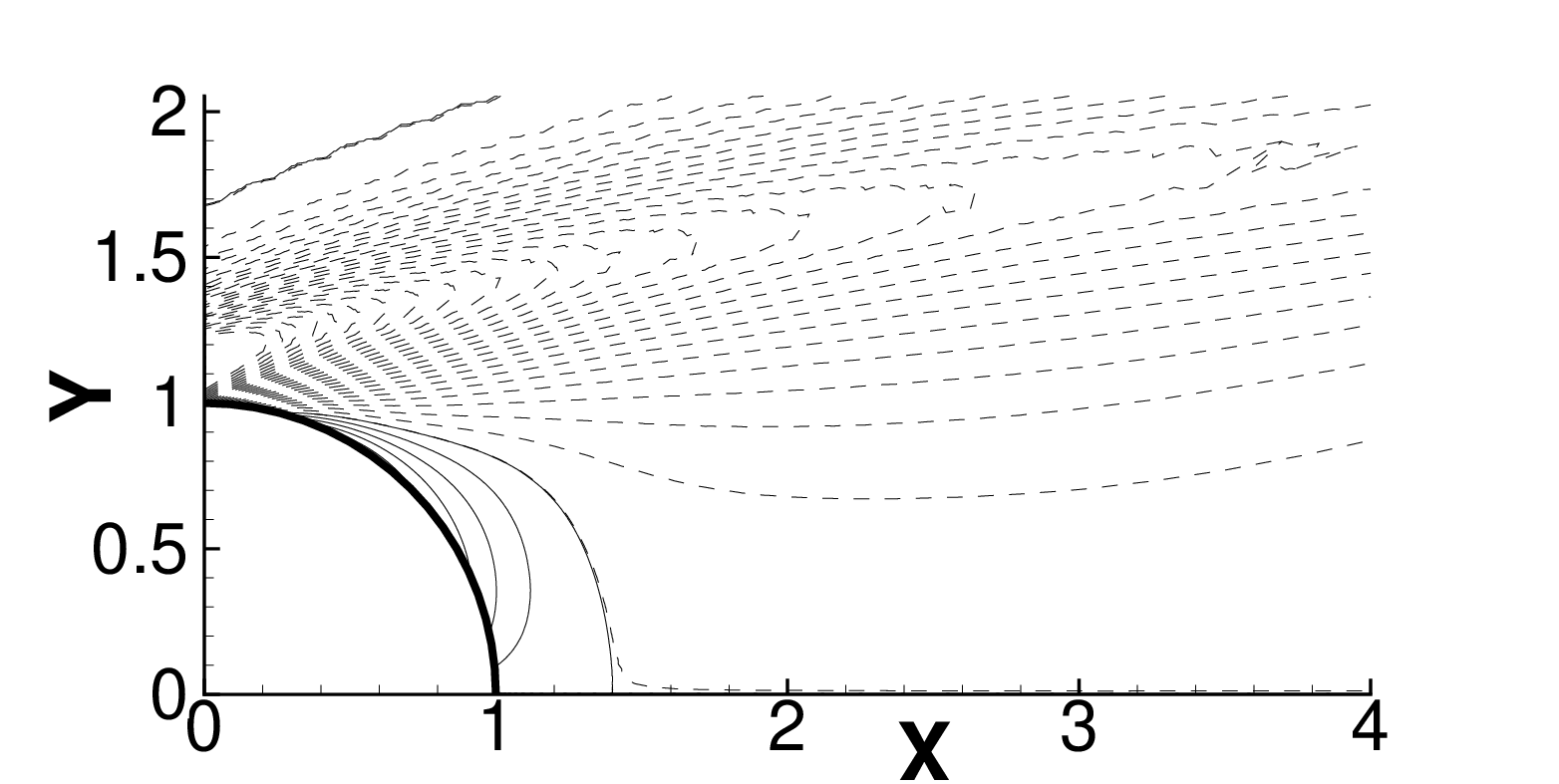}}}
\caption[]{(Continued, see previous caption for details)}
\end{figure}

\begin{figure}\centering{
\includegraphics[width=1.0\textwidth]{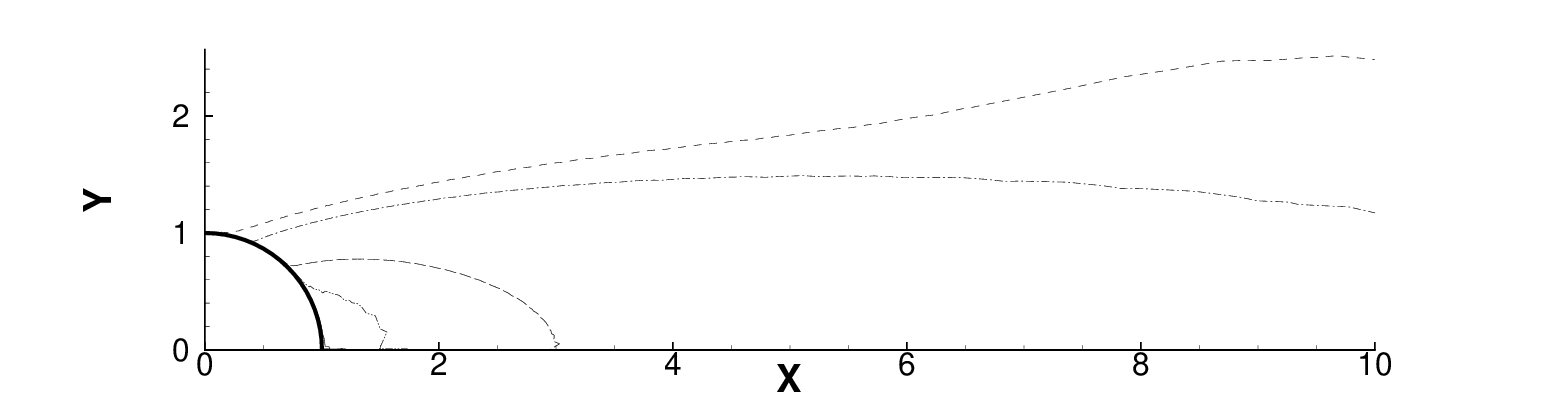}}
\caption{Boundaries of the separated regions in steady Navier-Stokes
  flows at the Reynolds numbers $Re = 10, 20, 100, 200$ (bigger
  recirculation bubbles correspond to higher Reynolds numbers).
  \label{fig:recirc}}
\end{figure}

\begin{figure}\centering{
\includegraphics[width=1.0\textwidth]{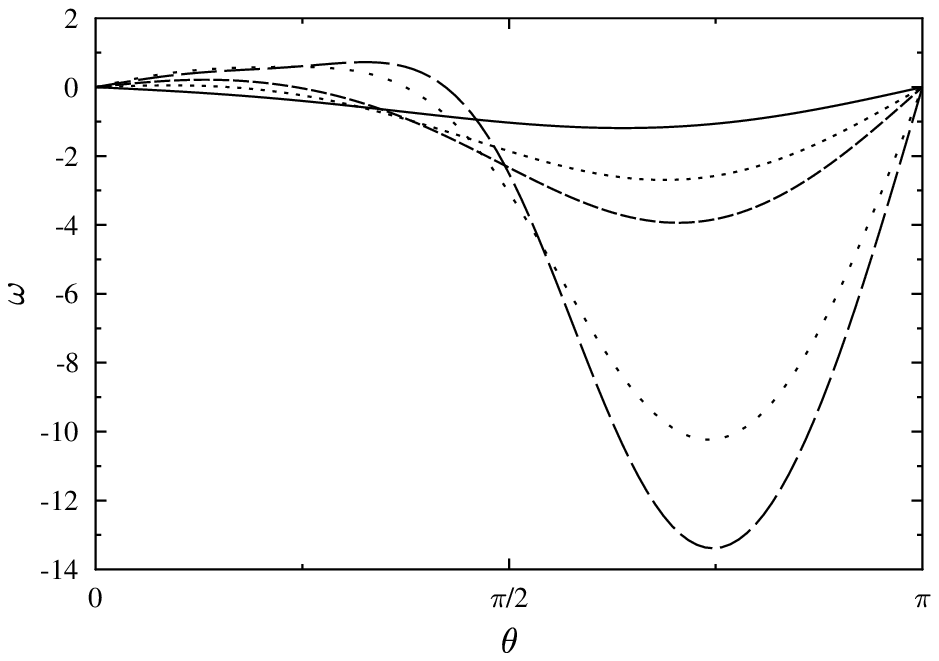}}
\caption{Vorticity $\omega(1,\theta)$ on the surface of the cylinder
  $A$ as a function of angle $\theta$ in steady Navier-Stokes flows
  at the Reynolds numbers $Re = 2$, $10$, $20$, $100$, $200$ (larger
  vorticity magnitudes correspond to higher Reynolds numbers).  }
\label{fig:vorticityonthesurface}
\end{figure} 

\begin{figure}
\centering \mbox{
\subfigure[$Re=10$]{
\includegraphics[width=0.48\textwidth]{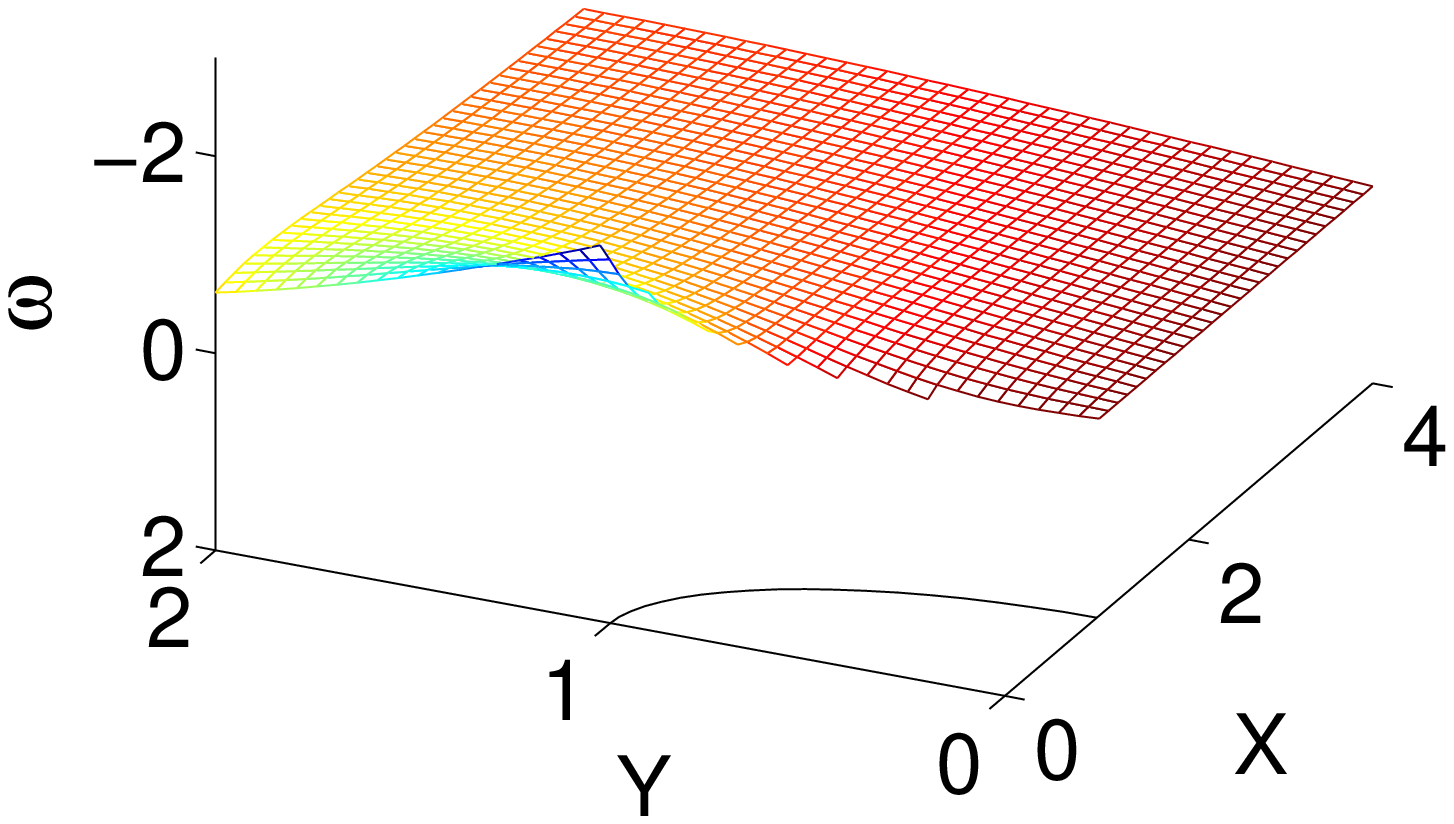}\label{fig:surfacevorticityRe5}}
\subfigure[$Re=200$]{
\includegraphics[width=0.48\textwidth]{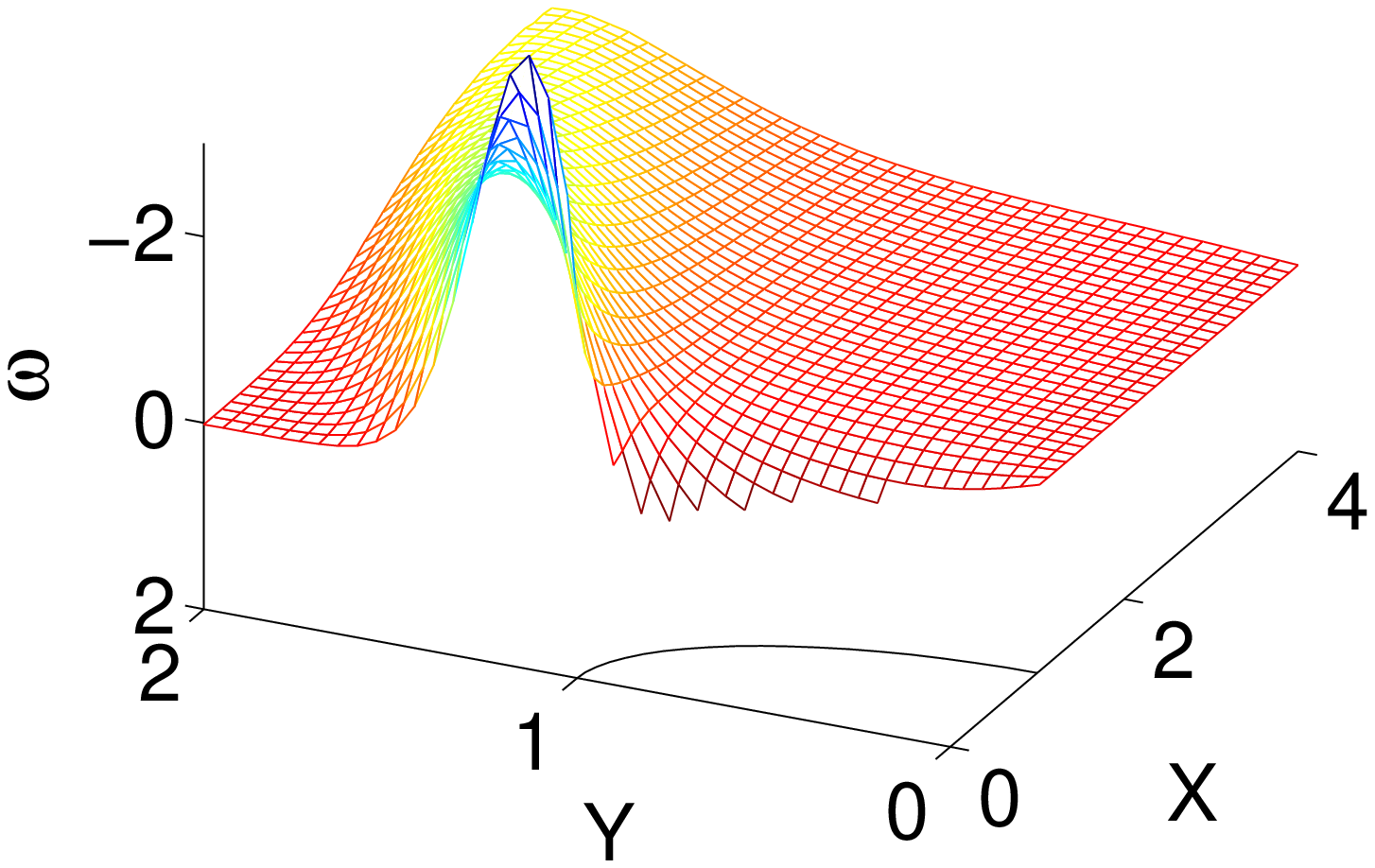}}\label{fig:surfacevorticityRe100}}
\caption{Surface plots of vorticity $\omega(x,y)$ in the
  steady Navier-Stokes flows at the Reynolds numbers indicated. The
  boundary of the cylinder is marked with a solid line and the vertical axes are reversed.
  \label{fig:surfacevorticity}}
\end{figure}

\begin{figure}
\centering
\mbox{
\subfigure[$L_R$]{
\includegraphics[width=0.48\textwidth]{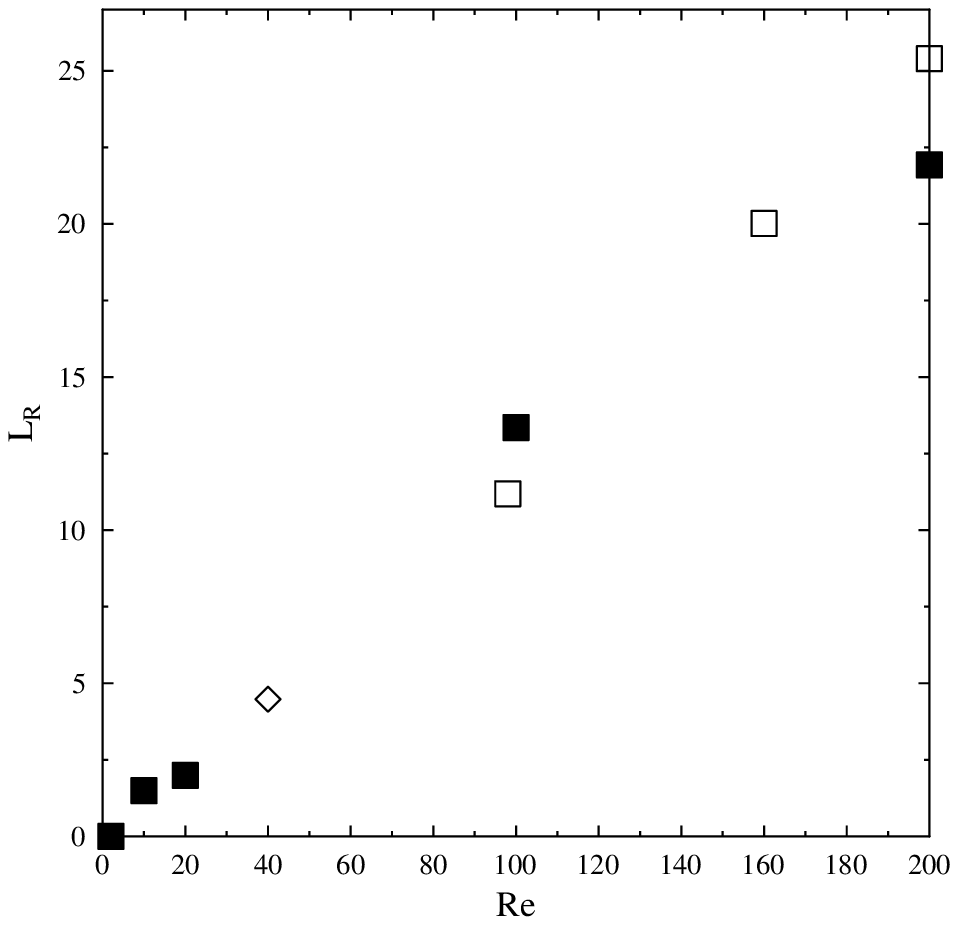}\label{fig:length}}
\subfigure[$W_R$]{
\includegraphics[width=0.48\textwidth]{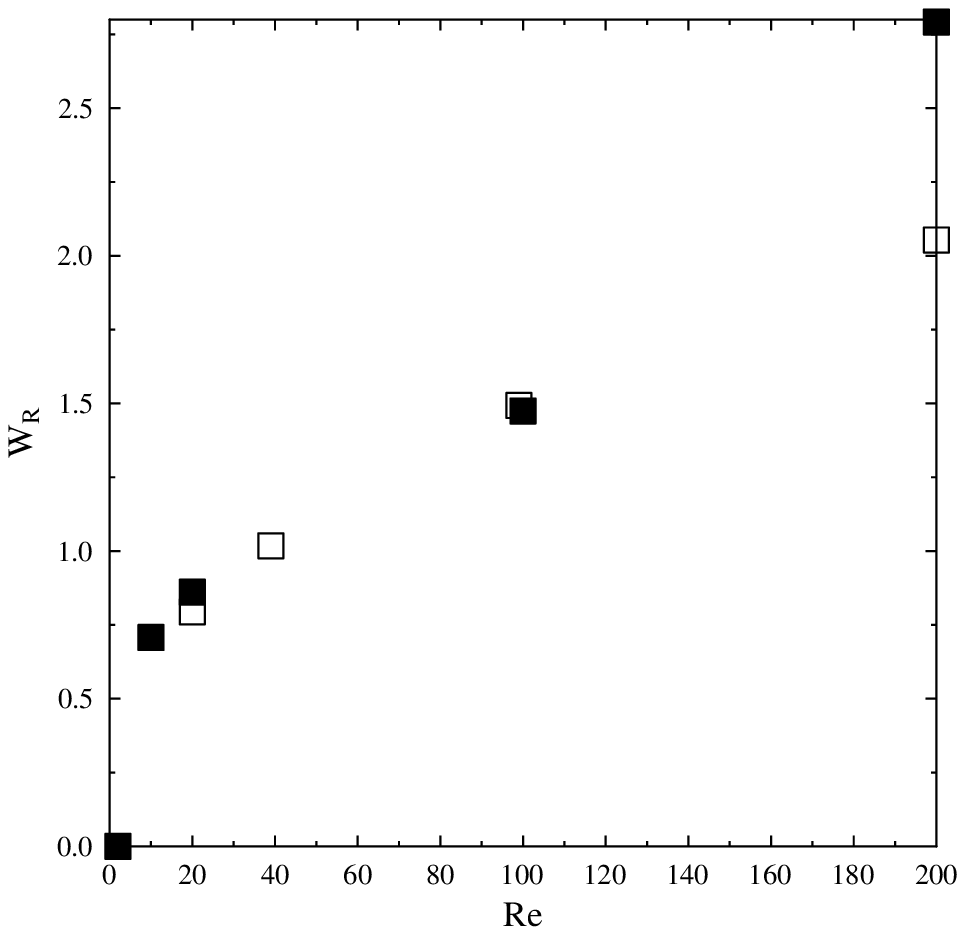}}\label{fig:width}}
\mbox{
\subfigure[$c_D$]{
\includegraphics[width=0.48\textwidth]{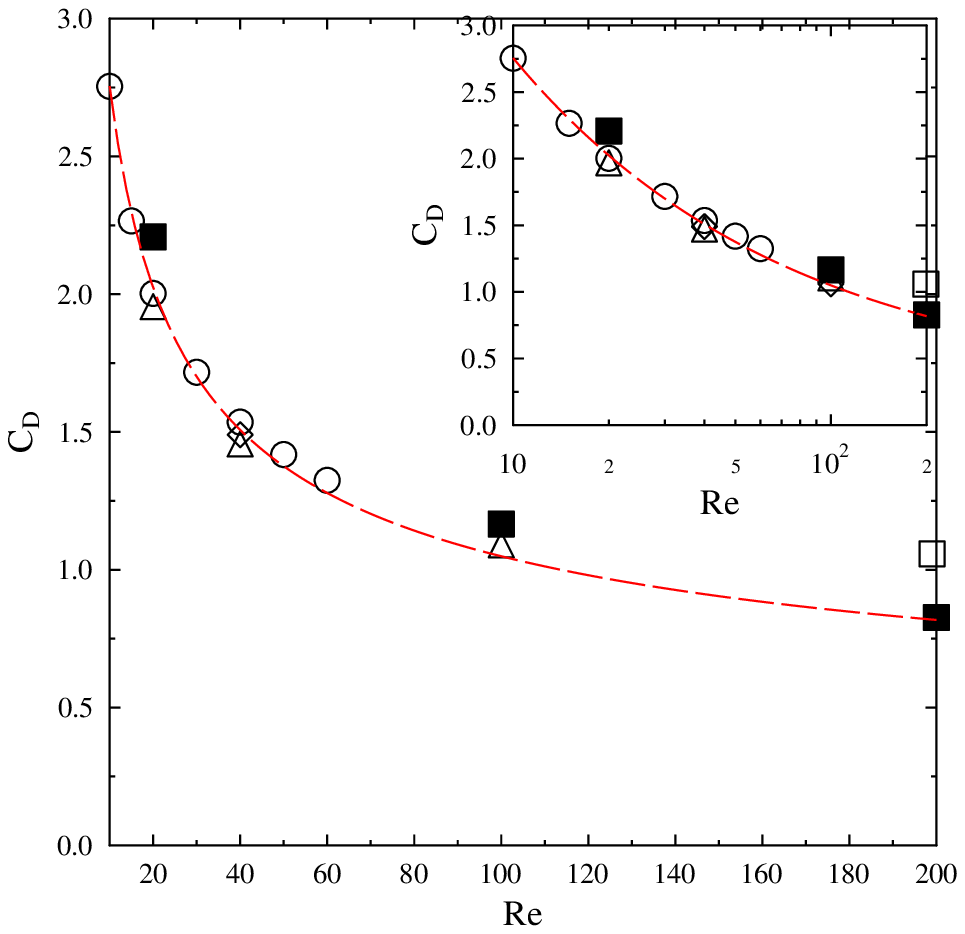}}\label{fig:CD}
\subfigure[$\theta_0$]{
\includegraphics[width=0.48\textwidth]{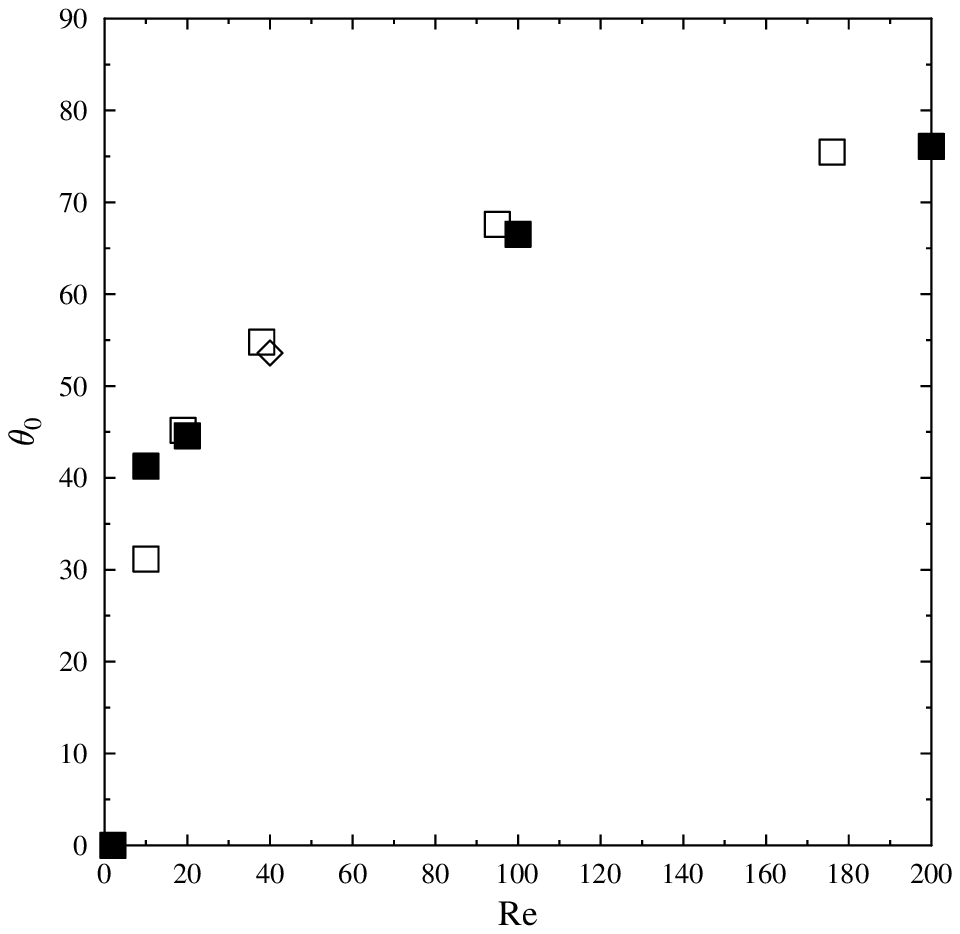}}\label{fig:sepangle}}
\caption{(a) Length $L_R$ and (b) half-width $W_R$ of the
  recirculation zone, (c) drag coefficient $c_D$ and (d) the
  separation angle $\theta_0$ as a function of the Reynolds number
  $Re$; in all figures, solid squares represent the present results,
  whereas empty squares correspond to the results of
  \cite{Fornberg1980,Fornberg1985} and diamonds correspond to the
  results of \cite{gbl13}. In Figure (c), empty circles correspond to
  the results of \cite{Nieuwstadt1973}, empty triangles correspond to
  the results of \cite{DAlessio1993} and dashed line represents the
  empirical fit obtained in \cite{smb09}. The inset in Figure (c)
  shows the same data as the main figure, but with the logarithmic
  scaling of the horizontal axis.}
\label{fig:FDmisc}
\end{figure}

\section{Conclusions}
\label{sec:final}

In this study we developed and validated a computational approach to
the solution of the steady 2D Navier-Stokes system in unbounded
domains.  The proposed method ensures that the solutions are
constructed to satisfy a number of properties of such flows revealed
by mathematical analysis. They concern the behavior of the velocity
fields at large distances from the obstacle, more specifically, their
slow decay towards the limiting values (slower than in the
time-dependent case). Some related ideas were already discussed in
earlier studies \cite{Fornberg1980,Fornberg1985,Bonisch2006,bhw05} and
in recent investigation \cite{gbl13}. Our method relies on the use of
rational Chebyshev polynomials to discretize functions defined on
unbounded domains, thereby avoiding the need to truncate the domain to
a finite computational box. Formulation of the steady 2D Navier-Stokes
flows in terms of streamfunction and vorticity results in the former
field having a discontinuity at infinity. It is resolved by recasting
the problem in terms of a ``skeleton'', related to the corresponding
Oseen flow, and suitable perturbation fields. The use of spectral
discretization allows us to achieve good accuracy with modest
resolutions. While we were able to obtain solutions for the Reynolds
numbers spanning two orders of magnitude, computation of flows for $Re
\ge 200$ was made difficult by the slow convergence of iterations
\eqref{eq:iter}. Possible reasons include poor conditioning of the
algebraic system solved to determine Newton's direction $- \left[
  \bnabla\bF({\bf X}^n)\right]^{-1} {\bf F}({\bf X}^n)$ and the
coupling of the drag force $F_x$ to the states $\left[
  \boldsymbol{\hat{\psi}} \ \ \boldsymbol{\hat{\omega}} \right]^T$
during iterations, cf.~\eqref{eq:Fx}. Reliable computation of steady
flows for large Reynolds numbers to address the fundamental questions
mentioned in the Introduction remains therefore an objective for
future research. Other related open questions concern computation of
steady flows in three dimensions and flows possibly involving
asymmetric wakes.

\section*{Acknowledgements}

The authors acknowledge the financial support from SHARCNET through a
graduate scholarship and from the Natural Sciences and Engineering
Research Council of Canada (NSERC) in the form of a Discovery Grant.
Calculations were performed using the high-performance computing
facilities provided by SHARCNET.

\appendix
% this is intended to fix the problem with appendices in the table of content
\renewcommand*{\thesection}{\Alph{section}}

\section{Discrete Spectral Chebyshev Differentiation --- Collocation Approach}
\label{sec:chebdiff}

We describe below the construction of differentiation matrices
corresponding to the spectral Chebyshev differentiation in the
collection setting. In combination with chain-rule formulas
\eqref{eq:derivativer}, they allow us to represent the derivatives
with respect to the radial coordinate $r \in [1,\infty]$. The present
approach is standard \cite{Canuto2007} and is described here for
completeness. Given a function $h(\xi) : \I \rightarrow \RR$, the
derivatives can be approximated as
\begin{subequations}
\label{eq:dx}
\begin{align}
\frac{\partial h(\xi_i)}{\partial x} & \simeq \sum_{j=0}^{N_2-1} \bar{d}^{(1)}_{i,j}h(\xi_j),\quad i=0,\ldots,N_2-1, 
\label{eq:dxa} \\
\frac{\partial^2 h(\xi_i)}{\partial x^2} & \simeq \sum_{j=0}^{N_2-1} \bar{d}^{(2)}_{i,j}h(\xi_j), \quad i=0,\ldots,N_2-1,
\end{align}
\end{subequations}
where $\xi_i$ and $\xi_j$ are the collocation points defined in
\eqref{eq:collocation} and the entries of the first differentiation
matrix are \cite{Peyret2002}
\begin{equation}\label{eq:diffone}
\bar{d}^{(1)}_{i,j} = \frac{\bar{c}_i}{\bar{c}_j} \frac{(-1)^{i+j}}{\xi_i-\xi_j},\quad 0 \leq i,j < N_2, \quad i \neq j,
\end{equation}
where $\bar{c}_0 = \bar{c}_N = 2$, $\bar{c}_j = 1$ for $1 \leq j
< N_2$. The diagonal elements are calculated as proposed in
\cite{Bayliss1995}
\begin{equation}\label{eq:Bayliss}
\bar{d}^{(1)}_{i,i} = - \sum_{j=0, j \neq i}^{N_2-1}\bar{d}^{(1)}_{i,j}, \quad 0 \leq i < N_2,
\end{equation}
which is intended to improve the accuracy of the differentiation
matrices by making sure that the numerical derivative of a constant
function is equal to zero. Indeed, by setting $h(x) = \text{const}$ in
equation \eqref{eq:dxa}, it is clear that the derivative vanishes at
all collocations points. Discrete differentiation given in
\eqref{eq:dxa} can be expressed in terms of differentiation matrix ${\bf \bar{D}}^1$
\begin{equation}\label{eq:D1}
[{\bf \bar{D}}^1]_{i,j} = \bar{d}^{(1)}_{i,j}, \quad 0 \leq i,j < N_2.
\end{equation}
The second-derivative differentiation matrix is obtained as
\begin{equation}\label{eq:D2}
{\bf \bar{D}}^2 = {\bf \bar{D}}^1 \cdot {\bf \bar{D}}^1,
\end{equation}
where $\cdot$ denotes the matrix-matrix multiplication and the
diagonal entries are calculated in a similar way to
\eqref{eq:Bayliss}, i.e., 
\begin{equation}
\bar{d}^{(2)}_{i,i} = - \sum_{j=0, j \neq i}^{N_2-1}\bar{d}^{(2)}_{i,j}, \quad 0 \leq i < N_2.
\end{equation}

\section*{References}

\bibliographystyle{natbib}
%\bibliographystyle{elsarticle-num.bst}
%\bibliography{bibtex}

\newcommand{\SortNoop}[1]{}

\end{document}